\documentclass[utf8]{frontiersSCNS} % for Science, 
\usepackage{url,lineno,microtype,subcaption}
\usepackage{scalerel,tikz}
\usepackage[flushleft]{threeparttablex}
\usepackage{longtable}
\usepackage{amsmath}	% Advanced maths commands
\usepackage{amssymb}	% Extra maths symbols
\usepackage{color}
\usepackage{algorithm}
\usepackage{booktabs}
\usepackage[noend]{algpseudocode}
\usepackage{pifont}% http://ctan.org/pkg/pifont
\usepackage{soul}
\usepackage{float}
\usepackage{multibib}
\newcites{Appendix}{Appendix References}
\usepackage{natbib}
\usepackage[flushleft]{threeparttablex}
\usepackage{mathrsfs}
\usepackage{mathtools}
\usepackage{tikz}
\usepackage{esint}
\usepackage{mathtools, nccmath}
\usepackage[draft]{todonotes}
\usepackage[colorlinks=true,
          citecolor=blue,
          linkcolor=blue,
          urlcolor=blue,
          linktocpage=true,
          hyperfootnotes=false,
          breaklinks=true]{hyperref}

\usetikzlibrary{svg.path}
\definecolor{orcidlogocol}{HTML}{A6CE39}
\tikzset{orcidlogo/.pic={\fill[orcidlogocol] svg{M256,128c0,70.7-57.3,128-128,128C57.3,256,0,198.7,0,128C0,57.3,57.3,0,128,0C198.7,0,256,57.3,256,128z}; \fill[white] svg{M86.3,186.2H70.9V79.1h15.4v48.4V186.2z} svg{M108.9,79.1h41.6c39.6,0,57,28.3,57,53.6c0,27.5-21.5,53.6-56.8,53.6h-41.8V79.1z M124.3,172.4h24.5c34.9,0,42.9-26.5,42.9-39.7c0-21.5-13.7-39.7-43.7-39.7h-23.7V172.4z} svg{M88.7,56.8c0,5.5-4.5,10.1-10.1,10.1c-5.6,0-10.1-4.6-10.1-10.1c0-5.6,4.5-10.1,10.1-10.1C84.2,46.7,88.7,51.3,88.7,56.8z};}}
\newcommand\orcidicon[1]{\href{https://orcid.org/#1}{\mbox{\scalerel*{
\begin{tikzpicture}[yscale=-1,transform shape]\pic{orcidlogo};
\end{tikzpicture}}{|}}}}
\usepackage[onehalfspacing]{setspace}

\newcommand{\noop}[1]{}
\newcommand{\appropto}{\mathrel{\vcenter{
  \offinterlineskip\halign{\hfil$##$\cr
    \propto\cr\noalign{\kern2pt}\sim\cr\noalign{\kern-2pt}}}}}
\newcommand{\var}[1]{\sigma^2_{#1}}
\newcommand{\vecB}[1]{\mathbf{#1}}
\newcommand{\Exp}[1]{\left\langle #1 \right\rangle}

\newcommand{\cor}[1]{\ell_{\text{cor},#1}}
\newcommand{\ten}[1]{\times10^{#1}}

\renewcommand{\d}[1]{\ensuremath{\operatorname{d}\!{#1}}}

\DeclareMathOperator\V{\mathcal{V}}
\newcommand{\M}{\mathcal{M}}
\DeclareMathOperator\Ma{\mathcal{M}_{\text{A}}}
\DeclareMathOperator\Mao{\mathcal{M}_{\text{A0}}}

\newcommand{\MaO}[1]{\mathcal{M}^{#1}_{\text{A0}}}

\DeclareMathOperator\kapar{\kappa_{\parallel}}

\DeclareMathOperator\Bo{\vecB{B}_0}

\newcommand{\dB}{\delta\mathbf{B}}

\DeclareMathAlphabet\mathbfcal{OMS}{cmsy}{b}{n}

\newcommand{\lva}{\Tilde{v}_{\text{A}}}
\newcommand{\va}{v_{\text{A}}}
\newcommand{\vao}{v_{\text{A0}}}
\newcommand{\lvai}{{\Tilde{v}_{\text{A,ion}}}}
\newcommand{\vai}{{v_{\text{A,ion}}}}
\newcommand{\vaoi}{{v_{\text{A0,ion}}}}
\newcommand{\lB}{\Tilde{B}}

\defcitealias{Brunt2010a}{BFP2010}
\defcitealias{Beattie2020}{BF2020}
\defcitealias{Kolmogorov1941}{K41}
\defcitealias{Menon2020}{MFK2020}
\defcitealias{Hopkins2013}{H2013}
\defcitealias{Skalidis2021}{S+2021}

\graphicspath{{Figures/}}
\def\keyFont{\fontsize{8}{11}\helveticabold }
\def\firstAuthorLast{Beattie {et~al.} 2022} %use et al only if is more than 1 author
\def\Authors{James R. Beattie\,$^{\orcidicon{0000-0001-9199-7771}\,1,*}$, Mark R. Krumholz\,$^{\orcidicon{0000-0003-3893-854X}\,1,2}$, Christoph Federrath\,$^{\orcidicon{0000-0002-0706-2306}\,1,2}$, Matt L. Sampson$^{\orcidicon{0000-0001-5748-5393}\,1}$ and Roland M. Crocker$^{\orcidicon{0000-0002-2036-2426}\,1}$}
\def\Address{$^{1}$Research School of Astronomy and Astrophysics, Australian National University, Canberra, ACT 2611, Australia \\
$^{2}$Australian Research Council Centre of Excellence in All Sky Astrophysics (ASTRO3D), Canberra, ACT 2611, Australia}
\def\corrAuthor{James R. Beattie}
\def\corrEmail{james.beattie@anu.edu.au}

\begin{document}
\onecolumn
\firstpage{1}

\title[Ion Alfv\'en velocity fluctuations]{Ion Alfv\'en velocity fluctuations and implications for the diffusion of streaming cosmic rays} 

\author[\firstAuthorLast ]{\Authors} 
\address{} 
\correspondance{}
\extraAuth{}
\maketitle

\begin{abstract}
The interstellar medium (ISM) of star-forming galaxies is magnetized and turbulent. Cosmic rays (CRs) propagate through it, and those with energies from $\sim\,\rm{GeV} - \rm{TeV}$ are likely subject to the streaming instability, whereby the wave damping processes balances excitation of resonant ionic Alfv\'en waves by the CRs, reaching an equilibrium in which the propagation speed of the CRs is very close to the local ion Alfv\'en velocity. The transport of streaming CRs is therefore sensitive to ionic Alfv\'en velocity fluctuations. In this paper we systematically study these fluctuations using a large ensemble of compressible MHD turbulence simulations. We show that for sub-Alfv\'enic turbulence, as applies for a strongly magnetized ISM, the ionic Alfv\'en velocity probability density function (PDF) is determined solely by the density fluctuations from shocked gas forming parallel to the magnetic field, and we develop analytical models for the ionic Alfv\'en velocity PDF up to second moments. For super-Alfv\'enic turbulence, magnetic and density fluctuations are correlated in complex ways, and these correlations as well as contributions from the magnetic fluctuations sets the ionic Alfv\'en velocity PDF. We discuss the implications of these findings for underlying ``macroscopic" diffusion mechanisms in CRs undergoing the streaming instability, including modeling the macroscopic diffusion coefficient for the parallel transport in sub-Alfv\'enic plasmas. We also describe how, for highly-magnetized turbulent gas, the gas density PDF, and hence column density PDF, can be used to access information about ionic Alfv\'en velocity structure from observations of the magnetized ISM.

\tiny
\keyFont{ \section{Keywords:} Magnetic Fields, Magnetohydrodynamic Turbulence, Multi-Phase Interstellar Medium, Galaxies, Cosmic Rays} 
\end{abstract}

\section{Introduction}\label{sec:intro}
    Magnetized turbulence is the rule and not the exception for the dynamics of the interstellar medium (ISM) in star-forming galaxies. Turbulence is a high-Reynolds-number $(\mathrm{Re} > 10^3)$ fluid state, where the Reynolds number is defined as $\mathrm{Re} = (\sigma_V L) / \nu$, and $\sigma_V$ is the velocity dispersion on length scale $L$ with kinematic viscosity $\nu$. Most astrophysical systems are vastly larger than the scales that are important for viscosity, and hence, turbulence has spread across most scales in the galaxies, with typical star-forming, cold molecular clouds boasting $\mathrm{Re}\sim 10^9$ \citep{Krumholz2015}. Due to the turbulent dynamo \citep[e.g.,][]{schekochihin2004simulations,Xu2016_conducting_dynamo,Federrath2016_dynamo,McKee2020,Xu2021_ionised_dynamo,seta2021saturation}, the energy in magnetic fields of the ISM is roughly at equipartition with the turbulent kinetic energy \citep{Boulares1990_CR_equipartition,Zweibel1995_energy_equipartition,Beck_2013_Bfield_in_gal,Seta2019_CRequipartition}. Magnetic fields play a dynamical role in the ISM, acting as a scaffold for the gas density through large-scale flux-freezing, facilitating some of the rich structure that we observe, e.g., in ISM observations, \citep{Li2011,Li2013,Soler2013,Planck2016a,Planck2016b,Federrath2016_brick,Cox2016,Malinen2016,Tritsis2016,Soler2017,Tritsis2018b,Heyer2020,Pillai2020} and simulations of ISM turbulence \citep{Soler2017a,Tritsis2018b,Beattie2020,Seifried2020,Kortgen2020,Barreto2021}. The ISM is also energy dense in relativistic, charged particles -- cosmic rays. 

    \subsection{Cosmic rays and the streaming instability}
    Cosmic rays (CRs) are high-energy (non-thermal), charged particles, that spiral around magnetic field lines at a radius set by the balance between Lorentz and centrifugal forces. Averaged over the ISM in star-forming galaxies, the energy densities of CRs, turbulent motions, and magnetic fields are roughly in equipartition \citep{Boulares1990_CR_equipartition,Beck_2013_Bfield_in_gal,Seta2019_CRequipartition}. CRs are important for understanding the ionization  (hence chemistry) and thus heating of interstellar gas \citep{Field1969,Xu2013,Krumholz2020}, and in turn influence the evolution of galaxies. For example, CR pressure gradients can significantly impact the morphology of simulated, ideal galaxies \citep{Salem2014_CRsGalaxies} and can drive and sustain galactic winds (and more general outflows) with mass-loading factors of order unity, which in turn can excite turbulent gas motions \citep{Uhlig2012_CRwinds,Booth2013_CRwinds,Girichidis2016_CR_outflows,Crocker2021_CR_sf_galaxies_1,Crocker2021_CR_sf_galaxies_2}. Because these processes depend upon CR pressure gradients, transport of CRs through the ISM is important to understand.    
    
    If magnetic fields in the ISM were static and structureless on scales of the CR gyroradius, CR transport would be trivial -- CRs would simply spiral along field lines, moving down them at the speed of light times the cosine of the pitch angle between the CR velocity vector the local magnetic field. However, Alfv\'en waves with frequencies comparable to the frequency of CR gyration can resonantly scatter CRs, randomly changing their pitch angle. Moreover, when a population of CRs is numerous enough, they themselves can excite such scattering waves via the streaming instability \citep{Lerche1967_streaming, Kulsrud1969_streaming,Wentzel1969_streaming,Skilling1971_streaming}. 
    CRs that have an energy range between $\sim\,\rm{GeV} - \rm{TeV}$, which dominate the CR pressure budget \citep{Evoli}, are likely to excite waves so efficiently that to zeroth order the CRs are scattered isotropically in pitch angle around field lines, still traveling at relativistic velocities. However, to first order, a slight asymmetry in the scattering distribution develops such that there is a bulk velocity along the field, $v_{\rm stream}$, that approaches the ion Alfv\'en speed, $\vai = B/\sqrt{4\pi\chi\rho}$, where $B$ is the magnitude of the local magnetic field, $\rho$ is the gas density and $\chi$ is the ionization fraction by mass. The asymmetry corresponds to an advective process whereby the population of scattering cosmic rays move down CR pressure gradients and along the field lines \citep{bell2013cosmic,caprioli2009dynamical,Krumholz2020,Bustard2020CosmicMedium,Xu2021_CR_streaming,Sampson2022_inprep_SCR_diffusion}. Throughout this study, we will refer to CRs that have self-confined to stream at close to $\vai$ as SCRs (streaming cosmic rays). 
    
    As we have described, the streaming instability is an advective process, e.g., the small asymmetry in the scattering angle distribution leads to the population of SCRs being advected down pressure gradients, along field lines. However, consider now multiple populations of SCRs distributed across a plasma and that we are ``observing" on length scales larger than the correlation length scale\footnote{Here we provide a rough estimate of the correlation scale of the magnetic field, as pertaining to our estimate in this paragraph. Consider an Alfv\'en wave traveling along a magnetic field line, over some time, $t_{\rm nl}$ that sets the timescale for the Alfv\'en wave to decorrelate. Clearly, by causation, this is proportional to the parallel spatial correlation length of the magnetic field, $\ell_{\rm cor} \sim \va t_{\rm nl}(\ell_{\rm cor} / \ell_0)$. $t_{\rm nl}$ is the nonlinear timescale for a turbulent fluctuation to decorrelate in the plasma, $t_{\rm nl}(\ell/\ell_0) = \ell\sigma_V^{-1}\left( \frac{\ell}{\ell_0} \right)^{-p} = \ell^{1-p} \ell_0^{-p} \sigma_{V}^{-1}$, where $\sigma_V$ is the system-scale velocity dispersion and $\ell_0$ is the driving scale of the turbulence. $p$ encodes the turbulence model ($p = 1/3$, \citealt{Kolmogorov1941}; $p = 1/2$, \citealt{Burgers1948}; $p=1/4$, \citealt{Kraichnan1965_IKturb}). Hence, $\ell_{\rm cor} \sim \va\ell_{\rm cor}^{1-p} \ell_0^{p} \sigma_V^{-1}$, and $\ell_{\rm cor}/\ell_0 \sim \mathcal{M}_{\text{A}}^{-1/p}$.} of the magnetic field embedded in the medium, $\ell_{\rm cor}/\ell_0 \sim \mathcal{M}_{\text{A}}^{-2}$, where $\Ma = \sigma_V / \va$ is the Alfv\'en Mach number, assuming a supersonic, Burgers velocity dispersion - size relation, $\sigma_v(\ell/\ell_0) \sim (\ell/\ell_0)^{1/2}$ \citep{Federrath2021}. On these scales, the field is tangled, chaotic (at least for $\Ma \gtrsim 2$, \citealt{Beattie2020_mag_fluc}), and different regions in space have causally disconnected magnetic fields. For average ISM parameters, $\Ma \approx 2$ \citep{Gaesnsler_2011_trans_ISM,Krumholz2020,Liu2021_cor_scales,Seta2021b}, $\ell_0 \sim 100\,\rm{pc}$ 
    \citep[$\sim$ the Galactic disk scale-height;][]{Karlsson2013_turbulence_scale,Goncalves2014_driving_scale,Li14b,Krumholz2018_metallicity_SF}, then on scales beyond $\ell_{\rm cor} \sim 25\,\rm{pc}$ streaming cosmic rays, locked to magnetic field lines, take tangled, chaotic paths that resemble a random walk, leading to a spatial dispersion between cosmic rays originating from the same source. We call this process the ``macroscopic diffusion" of SCRs.  
    
    However, it is not only tangling of magnetic field lines above the correlation length that may be responsible for creating spatial dispersion between populations of SCRs. Because SCRs become self-confined to travel at $\vai \propto B/\sqrt{\chi \rho}$ inhomogeneities in ionic Alfv\'en wave speeds may arise in either the magnetic field \textit{or} the gas density and contribute to the macroscopic diffusion of SCRs. Gas density fluctuations can occur on scales much smaller than the correlation scale of the magnetic field \citep[e.g., ion fluctuations down to less than AU scales,  ][]{Armstrong1995_power_law}. A detailed experimental study of diffusion for streaming cosmic rays is beyond the scope of this study, and is presented with great detail in \citet{Sampson2022_inprep_SCR_diffusion}. In this study we explore the ionic Alfv\'en wave fluctuations in a compressible, magnetized turbulent medium across a broad range of plasma parameters describing the diffuse, warm atomic medium to dense \citep[possibly sub-Alfv\'enic --][]{HuaBai2013,Federrath2016_brick,Hu2019,Heyer2020,Hwang2021,Hoang2021,Skalidis2021_obs_sub_alf} highly-supersonic molecular clouds, highlighting the role of compressibility in the generation of the inhomogeneities in ionic Alfv\'en wave speeds. To the authors' knowledge, these fluctuations have not been studied in the astrophysical compressible turbulence literature, even though Alfv\'en waves, and the speed in which they travel, have deep roots in MHD turbulence theory and phenomenology \citep[e.g.,][]{Elsasser1950_vars,Iroshnikov_1965_IK_turb,Kraichnan1965_IKturb,Sridhar1994_weak_turbulence,Goldreich1995,Boldyrev2006}. However, the ingredients for developing a theory for understanding the variance of ionic Alfv\'en velocities, and more broadly their 1-point volume-weighted PDF, have been developed. These are, of course, the density and magnetic field fluctuations. Turbulence theory is a collection of two-point statistical models; however in this study, we explore and develop $2^{\rm nd}$ moment theories, which are more easily applied to observations of interstellar gas. Before progressing to the results of this study we consider three theoretical aspects of the problem: (1) what is the state of the ionization by mass in typical ISM conditions? and  what is the nature of (2) the density and (3) the magnetic field fluctuations, up to $2^{\rm nd}$ moments, in compressible MHD turbulence? We start with the first.
    
    \subsection{On the ionization state of turbulent density fluctuations}\label{sec:ionization _state}
    
    We first consider the ionization fraction $\chi$, for the purposes of demonstrating why we need not account for its fluctuations separately. In equilibrium in a region with a constant ionization rate per neutral particle $\zeta$, the condition for equilibrium is simply balance between the ionization and recombination rates per unit volume,
    \begin{equation}
        \zeta n_n = \alpha_{\rm rec} n_e n_{\rm ion},
    \end{equation}
    where $n_n$, $n_e$, and $n_{\rm ion}$ are the number densities of neutral species, electrons, and ions, respectively and $\alpha_{\rm rec}$ is the recombination rate coefficient for free electrons with ions. For simplicity consider a region of weakly ionized plasma, $\chi\ll 1$, where all ions are singly ionized. In this case we have $n_{\rm ion} = n_e = \chi \rho / \mu_{\rm ion} m_{\rm H}$, where $\mu_{\rm ion}$ is the mean atomic mass of ions and $m_{\rm H}$ is the hydrogen mass. Similarly, we have $n_n = \rho/\mu m_{\rm H}$, where $\mu$ is the mean atomic mass of neutrals, and therefore 
    \begin{equation}\label{eq:ion_equilibrium}
        \chi = \left(\frac{\zeta \mu_{\rm ion}^2 m_{\rm H}}{\alpha_{\rm rec} \mu \rho}\right)^{1/2}.
    \end{equation}
    Thus in equilibrium we should expect $\chi \propto \rho^{-1/2}$.
    
    However, since we are interested in fluctuations, we must next ask whether equilibrium is an appropriate assumption. The timescale required for a given parcel of gas to reach ionization equilibrium is the ion density divided by the rate at which the ion density changes,
    \begin{equation}
        t_{\rm ion} = \frac{n_{\rm ion}}{\zeta n_n} = \frac{\chi \mu}{\zeta \mu_{\rm ion}}.
    \end{equation}
    Significantly, this does not depend on the density, except indirectly through $\chi$. We can therefore immediately determine characteristic values of $t_{\rm ion}$ for different phases of the ISM. In the atomic ISM, we generally expect to have $\chi \sim 10^{-3} - 10^{-1}$, $\mu = 1.4$ (for the standard cosmic mix of H and He), $\mu_{\rm ion} = 1$ (H is the dominant ionized species), and $\zeta \sim 10^{-16}$ s$^{-1}$ \citep{Wolfire2003}, and therefore $t_{\rm ion} \sim 0.4 - 40$ Myr; in the interior of a molecular cloud, the equilibrium ionization fraction is lower, $\chi \sim 10^{-6}$, and we have $\mu = 2.33$ (H$_2$ + He composition), $\mu_i = 29$ (HCO$^+$ is the dominant charge carrier -- \citealt{Krumholz2020}), and $\zeta \sim 10^{-16} - 10^{-17}$ s$^{-1}$, and $t_{\rm ion} \sim 30-300$ yr.
    
    This should be compared to the characteristic timescale over which the density changes which, for a turbulent medium, \citet{Scannapieco2018} show is given approximately by
    \begin{equation}
        t_{\rho/\rho_0} \approx \frac{\tau}{3} \left[\frac{1}{2} - \frac{1}{\pi} \arctan\left(\frac{s - s_*}{2}\right)\right],
    \end{equation}
    where $\tau$ is the the flow crossing time, $s=\ln(\rho/\rho_0)$ is the logarithmic over-density in a region with mean density $\rho_0$, and $s_*$ is a constant of order unity that depends on the Mach number and Alfv\'en Mach number of the flow. This timescale varies from $\tau/3$ to zero slowly as a function of $s$.
    
    The above allows a few immediate conclusions. In the molecular ISM, we can safely assume instantaneous equilibrium: molecular clouds have flow crossing times of $\sim 1$ Myr, compared to times of at most a few hundred years to reach ionization equilibrium. Even in the colder parts of the atomic ISM, which tend to have $\chi\sim 10^{-3}$, instantaneous equilibrium is probably a safe assumption, since characteristic crossing times of the atomic ISM are of order 10 Myr, while at $\chi\sim 10^{-3}$ we have $t_{\rm ion}\lesssim 1$ Myr. Only in the diffuse atomic ISM are we likely to encounter regions where the ionization equilibration time is comparable to or longer than the flow crossing timescale. For this reason, we will simply assume a relationship $\chi\propto \rho^{-1/p}$, for any $p$ in what follows. Clearly, $p\rightarrow\infty$ $(\chi\sim\mathrm{constant})$, corresponds to the diffuse atomic ISM, and $p = 2$ for the other phases, which are in ionization  equilibrium. Reality should lie between these two limiting cases. Having established the ionization  state of the density fluctuations, we now turn our attention to the spatial statistics of the density fluctuations themselves in the lognormal framework. 

    \subsection{Lognormal density fluctuation theory}\label{sec:intro:dens}
    
    One of the key differences between incompressible and compressible turbulence is the dynamical role of density fluctuations and shocked gas in the turbulent plasma. Lognormal models for the PDF of turbulent $\rho / \rho_0$ fluctuations first originate from \citet{Vazquez1994}. \citeauthor{Vazquez1994} considers a linear density fluctuation in a self-similar (scale-free), isothermal plasma, where thermal pressure, $(1/\M^2)\nabla c_s^2 \rho$, is negligible $(\M \gg 1)$, where $\M = \sigma_v / c_s$ is the sonic Mach number, and the gas is not bounded by self-gravity, $\alpha_{\rm vir} \propto E_{\rm kinetic}/|E_{\rm grav}| \gg 1$, where $ E_{\rm kinetic}$ is the kinetic and $|E_{\rm grav}|$ is the gravitational energy, respectively. With these assumptions in hand, the lognormal PDF is motivated by assuming that for time, $t_n$, a density fluctuation can be expressed as a multiplicative interaction through density fluctuations at previous times,
    \begin{align}\label{eq:fluctuation_chain}
        \frac{\rho(t_n)}{\rho_0} = \left(\frac{\rho(t_{n-1})}{\rho_0}\right) \left(\frac{\rho(t_{n-2})}{\rho_0}\right) \left(\frac{\rho(t_{n-3})}{\rho_0}\right) \hdots \left(\frac{\rho(t_{1})}{\rho_0}\right) \frac{\rho(t_0)}{\rho_0} = \left(\prod^{n-1}_{i=1} \frac{\rho(t_{i})}{\rho_0}\right) \frac{\rho(t_0)}{\rho_0},
    \end{align}
    where $\rho(t_0)/\rho_0 = \rho_0/\rho_0 = 1$ is the initial density, before the turbulent interactions, in units of the mean. Under the log-transformation, the density fluctuations become additive,
    \begin{align}
        s(t_n) = \ln\left[ \frac{\rho(t_n)}{\rho_0} \right] = \sum^{n-1}_{i=1} \ln\left[\frac{\rho(t_{i})}{\rho_0}\right] = \sum^{n-1}_{i=1} s_i,
    \end{align}    
    turning the problem into one that involves the sum of random variables. If each $\rho(t_n) / \rho_0\; \forall n$, is generated by the same underlying distribution and is statistically independent from each of the other fluctuations, $\Exp{s(t_n)s(t_m)}_t=\delta(t_n - t_m)$, i.e., each addition of an extra fluctuation to \autoref{eq:fluctuation_chain} does not depend upon the current state of $\rho(t_n) / \rho_0$), then  $\rho(t_n) / \rho_0\; \forall n$ are said to be independent, identically distributed variables, and we can apply the central limit theorem. As $n$ approaches infinity the central limit theorem states that the distribution of $s(t_n)$ is normal, specifically the distribution has the functional form,
    \begin{align}
        p_s(s | \sigma_s^2) &= \frac{1}{\sqrt{2\pi\sigma_s^2}} \exp \left\{ -\frac{(s- s_0)^2}{2\sigma_s^2} \right\}, \label{eq:dis} \\
        s &\equiv \ln(\rho / \rho_0), \\
        s_0 &= - \frac{\sigma_s^2}{2},
    \end{align}
    where $p_s(s;\sigma_s^2)$ is the probability distribution for $s$, with mean $s_0$ and variance $\sigma_s^2$. The variance of $s$ solely determines the distribution, which is a mathematical feature of the lognormal distribution. In principle, this means (assuming no spatial correlations) that any fluctuation that has had a long history of interactions is lognormally distributed. However, because \citet{Vazquez1994} assumed that the fluid is perfectly scale-free (i.e., invariant under arbitrary length scalings), the local fluctuation theory can be applied globally, on any scale, and hence the $s$-PDF ought to be normal at any scale on which we probe it\footnote{Note that this cannot be strictly true, as discussed in detail in \citet{Hopkins2013}, \citet{Squire2017}, and \citet{Beattie2021_spdf}, since the PDF on each scale is constructed from convolutions of PDFs from scales below it, and convolutions of lognormal PDFs do not result in lognormal PDFs. For this reason, assuming that PDFs on all scales are lognormal violates mass conservation, and recent extremely high resolution turbulence simulations with $> 10,000^3$ grid elements show clear variation of the PDF and the morphology with scale \citep{Federrath2021}. However, these are theoretical details, and empirically and practically speaking, the $s$-PDF is approximately normal. While some authors have recently provided extended models that address some of the shortcomings of the purely lognormal phenomenology \citep{Hopkins2013,Squire2017,Mocz2018}, we will not discuss them in detail in this study. For a summary of many of the theoretical works see \citet{Beattie2021_spdf}.}.  
    
    Because lognormal models of the $s$-PDF are solely parameterised by the $\sigma_s^2$, understanding the variance in supersonic and magnetized ISM turbulence has been of great interest\footnote{Note that even though we are focusing strictly on ISM turbulence, recent works have developed density variance relations for compressible, subsonic, stratified turbulence, with applications for understanding the nature of fluctuations in the intracluster medium \citep{Mohapatra2020,Mohapatra2020b}.}. \citet{Padoan1997_imf} showed that the density variance could be related to the typical shock-jump conditions in the plasma, which has led to a plethora of relations. In general, 
    \begin{align}
        \sigma^2_s &= f(\M,\Mao,b,\gamma,\Gamma), \label{eq:var}
    \end{align}
    is a function of the sonic Mach number \citep{Vazquez1994,Padoan1997_imf,Passot1998,Price2011,Konstandin2012a}, the Alfv\'en Mach number and the strength of the large-scale field \citep{Padoan2011,Molina2012_dens_var,Beattie2021_multishock,Beattie2021_spdf}, the turbulent driving parameter $b$ (the mixture of solendoial and compressive modes in the driving source) \citep{Federrath2008,Federrath2010}, and the thermodynamics, including the adiabatic index $\gamma$ \citep{Nolan2015} and the polytropic index $\Gamma$ \citep{Federrath2015_polytropic}. We will find that $\sigma_s^2$ plays a central role in determining the Alfv\'en velocity variance, adding yet another application case and reason for understanding and modeling the density fluctuations in compressible turbulence. In fact we will show that the density fluctuations completely govern the Alfv\'en velocity fluctuations, which has significant implications for compressible MHD phenomenology. However, now we turn our attention to the magnetic field. 
    
    \subsection{The fluctuating and large-scale magnetic field in compressible plasmas}\label{sec:intro:mag}
    
    Next consider the statistics of the magnetic field. Due to the small-scale dynamo action, magnetic field fluctuations that are roughly at equipartition \citep[e.g., ][]{Xu2016_conducting_dynamo,McKee2020,seta2021saturation} with the turbulent fluctuations are ubiquitous in both incompressible and compressible MHD turbulence across the Universe \citep{Beck_2013_Bfield_in_gal,Subramanian2016_origins_review,Subramanian2019_origins}. Once saturation has occurred, based on a balance between the magnetic and kinetic energies, in an isothermal supersonic plasma, $\Exp{\delta B^2}^{1/2}$ scales with $\M f(\Mao)$, where $f(\Mao)=c_s\sqrt{\pi\rho_0}\Mao$ for turbulence with a sub-Alfv\'enic large-scale field and $f(\Mao)=2c_s\sqrt{\pi\rho_0}\MaO{-1/3}$ in the super-Alfv\'enic regime  \citep{Federrath2016_dynamo,Beattie2020_mag_fluc,Beattie2022_inprep_energybalance}. This means that $\Exp{\delta B^2}^{1/2}/B_0 = \MaO{2}/2$ for sub-Alfv\'enic turbulence, or likewise $\Exp{\delta B^2}^{1/2} \propto B_0^{-1}$. This naturally leads to the question, is supersonic, sub-Alfv\'enic turbulence Alfv\'enic, in the sense that the nonlinear interactions are solely determined by counterpropagating Alfv\'enic fluctuations traveling along field lines  \citep[e.g.,][]{Elsasser1950_vars,Iroshnikov_1965_IK_turb,Kraichnan1965_IKturb}? Because supersonic $\Mao < 1$ turbulence is magnetically-dominated, it naively may seem Alfv\'enic, but as we will show, at least on large-scales (scales where $\sigma_v >c_s$) it is the density fluctuations that completely determine the Alfv\'en velocity fluctuations (and hence Alfv\'enic component of the turbulence) which is certainly not part of the regular Alfv\'enic turbulence phenomenology.
    
    %, and both the density and velocity fields are filled with strong shocks, so at least on large-scales, where the velocity dispersion is supersonic, the sub-Alfv\'enic turbulence in our study is not necessarily weak.
    
    % Fluctuations perpendicular to $\Bo$, namely Alfv\'enic fluctuations, are the messengers of $\beta < 1$ turbulent plasmas, where $\beta = 2 c_s^2 / \va^2$. This is because Alfv\'en wave fluctuations travel at $\va$, which is larger than $c_s$ for $\beta < 1$, and so these waves become an efficient way of exchanging information in the plasma, whether it be informing different parts of the magnetic field line that it is bending, or reacting to and resonating with an appropriately energized cosmic ray. Alfv\'en waves also play a vital role in incompressible turbulence phenomenology, where weak 4-wave interactions grow in strength until an emergence of a strong energy cascade that is critically balanced between the nonlinear timescale that it takes energy to cascade to smaller scales and the linear Alfv\'en wave period timescale \citep{Goldreich1995,Schekochihin2020_bias_review}. 
    
    In contrast to Alfv\'enic turbulence, which is determined by weak Alfv\'en and slow wave interactions or strong nonlinear critically balanced cascades \citep{Goldreich1995,Lithwick2001_compressibleMHD,Schekochihin2007_alfvenic_turb}, magnetic fluctuations parallel to $\Bo$ seem to play an important role in compressible sub-Alfv\'enic large-scale field turbulence \citep{Beattie2020_mag_fluc,Skalidis2020,Skalidis2021,Beattie2021_spdf,Beattie2022_inprep_energybalance}, which, as we have discussed in \autoref{sec:intro}, is relevant to cold molecular gas in the ISM \citep{HuaBai2013,Federrath2016_brick,Hu2019,Heyer2020,Hwang2021,Hoang2021,Skalidis2021_obs_sub_alf}. In Alfv\'enic turbulence, the parallel fluctuations passively trace slow modes that form along the magnetic field \citep{Goldreich1995,Lithwick2001_compressibleMHD,Schekochihin2009_cascades}, but in supersonic turbulence the parallel fluctuations are excited around sites of strong shocks along the large-scale field \citep[see Figure~10 in ][]{Beattie2021_spdf}. They also play a vital role in the formation of large-scale, non-turbulent structures in the plasma (which may be related to the formation of 2D condensates previously observed in incompressible plasmas, e.g., \citealt{Boldyrev2009_spectrum_of_weak_turb,Wang2011_residual_energy}). By decomposing the turbulence into linear modes and phases, \citet{Yang2019_structures_and_modes} found that $\approx$77\% of the total energy was in non-propagating structures (those that do not follow a theoretical wave dispersion relation from linear theory)\footnote{Note that this means that even though the turbulence is driven in the weak regime, it is not clear if the turbulence could faithfully be classified as weak if over 70\% of the energy budget in the fluid is from $k_{\parallel}=0$ structures that are not waves \citep{Boldyrev2009_spectrum_of_weak_turb,Yang2019_structures_and_modes}.}. These are system-scale $(k_{\parallel} = 0)$ rigid body vortices that, to become stationary, require parallel magnetic field pressure gradients (and hence strong parallel fluctuations that oppose $\Bo$) to balance the centrifugal force of the rotating fluid \citep{Beattie2020_mag_fluc,Beattie2021_spdf}. Because of this coupling between the vortex motions perpendicular to $\Bo$ and the shocked gas parallel to $\Bo$, they contain roughly an order of magnitude more energy than their perpendicular (Alfv\'enic) counterparts when the plasma is very sub-Alfv\'enic (but the same energy when the turbulence is super-Alfv\'enic) and in a quasi-stationary turbulence state act to balance the kinetic energy in these highly-compressible and sub-Alfv\'enic plasmas \citep[see Figure~B1 in ][]{Beattie2022_inprep_energybalance}. \citep{Beattie2022_inprep_energybalance}. We leave a detailed analysis of these vortices to a future study. With that, we have the required background to construct magnetic field variance relations based on energy balance, and we leave further discussion of the magnetic field fluctuations until we compare theory directly with our simulation results in \autoref{sec:magnetic}.
    
    %As well as naturally inducing large-scale (at the system scale) anisotropy on the plasma, the large-scale field actively influences the very nature of the magnetized turbulence. For example, the rms turbulent field follows a power law with the large-scale field strength $\Exp{\delta B^2}^{1/2} \propto B_0^{-1}$, and hence the turbulence with respect to the fluctuating magnetic field component is extremely super-Alfv\'enic when the large-scale field is strong 
    
    % This is a natural repercussion of $\Bo$ reducing the nonlinear terms (responsible for the turbulence) in the MHD fluid equations, and amplifying the linear terms \citep{oughton_priest_matthaeus_1994}. In the incompressible limit, and when $B_0 \gg \delta B$, based on fluid theory and numerical simulations, it is likely that the plasma motions become completely two-dimensional, reduced to decoupled planes of vortical fluid motions around $\Bo$ \citep{oughton_priest_matthaeus_1994,Alexakis2011,Verma2017,Beattie2020_mag_fluc}. 
    
    \subsection{Outline}
    
    This study is organized as follows: In \autoref{sec:sims} we introduce the isothermal, compressible ideal MHD models and simulation setup that we use to understand the Alfv\'en velocity fluctuations over a broad range of plasma parameters. In \autoref{sec:va_fluctuations} we begin our construction of the $\va$-PDF and variance by studying the density and magnetic field fluctuations, including the covariance between the two quantities. We include comparisons between compressible MHD turbulence theory discussed in this section, and the simulation data from our turbulence experiments. Then we develop a variance and 1-point volume-weighted PDF theory for the Alfv\'en velocity fluctuations for sub-Alfv\'enic MHD turbulence, where the fluctuations are dominated by the effects of compressibility. In \autoref{sec:discussion} we discuss the implications of Alfv\'en velocity fluctuations for column density observations of molecular clouds and macroscopic diffusion of cosmic rays undergoing the streaming instability in sub-Alfv\'enic regions of the ISM. Finally, in \autoref{sec:conclusion} we summarise and itemize the key results of the study. We list the unique mathematical notation and symbols that we use in this study in \autoref{tb:symbol_glossary}.
    
    \begin{figure*}
        \centering
        \includegraphics[width=0.84\linewidth]{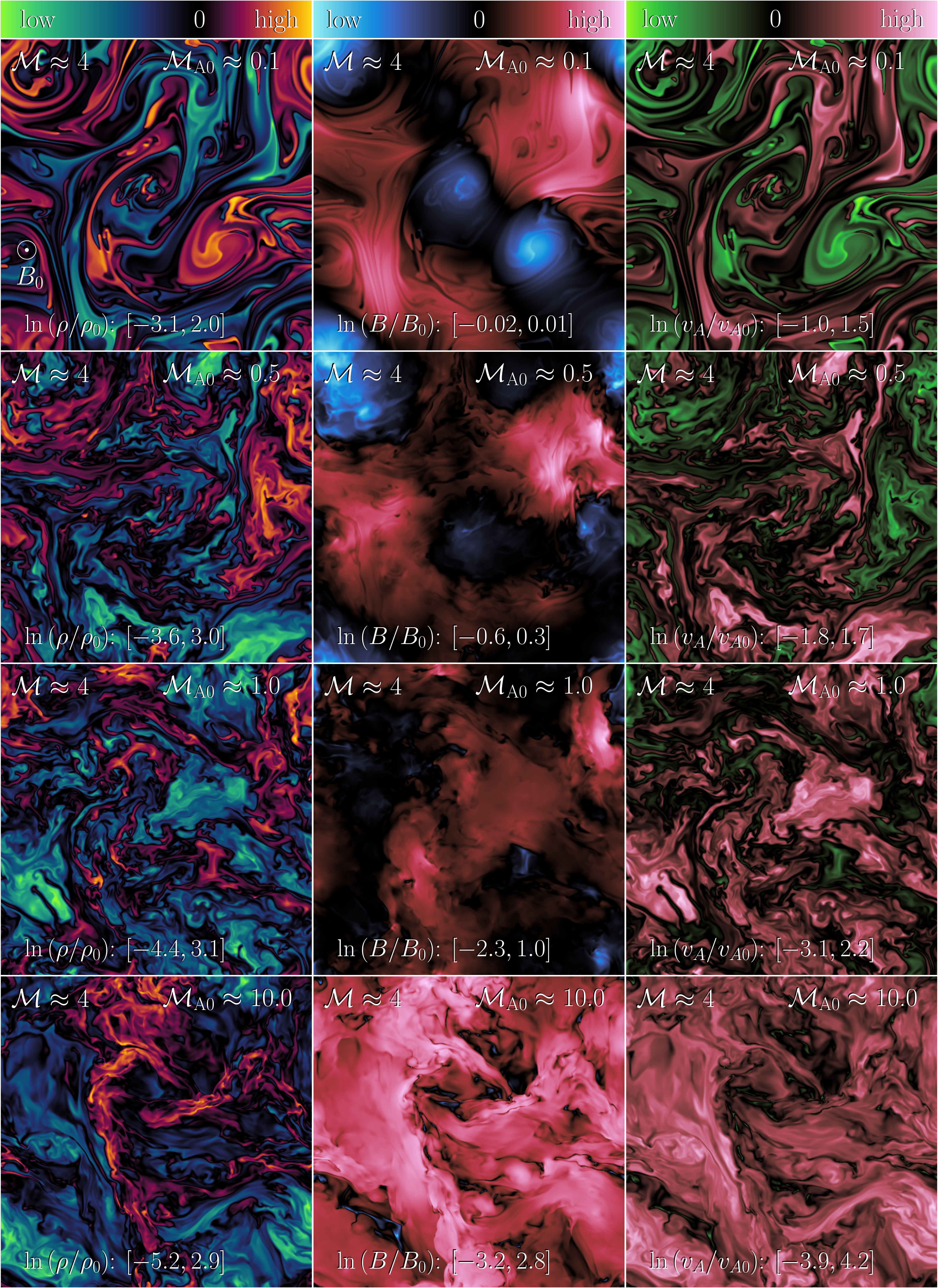}
        \caption{Two-dimensional slices through the logarithmic gas density, $\ln(\rho/\rho_0)$ (first column), logarithmic magnetic field, $\ln(B/B_0)$ (center column) and logarithmic Alfv\'en velocity magnitude, $\ln(\va/\vao)$ (third column) for $\M=4$ simulations. For each panel shown, simulation Mach and Alfv\'en Mach numbers are listed at the top, and the range covered by the color bar is indicated in brackets at the bottom. The slices are taken perpendicular to the large-scale field, $\vecB{B}_0$, which is pointing out of slice plane, as indicated in the top-left panel. The first three rows highlight the spatial correlation between the three field variables for the sub-to-trans-Alfv\'enic regime $(\Mao \lesssim 1)$, and the last row for the super-Alfv\'enic regime $(\Mao > 1)$. For the sub-to-trans-Alfv\'enic, the spatial structure in $\ln(\va/\vao) \sim \ln(\rho/\rho_0)$, but as $\Mao$ grows $\ln(\va/\vao)$ begins to more closely match $\ln(B/B_0)$.}
        \label{fig:12_panel_gas_vars}
    \end{figure*}

\section{Turbulence simulations}\label{sec:sims}

    \subsection{Stochastically driven isothermal MHD fluid model}
    To understand the nature of Alfv\'en velocity fluctuations in MHD turbulence, we use a modified \citep[see Methods section in ][]{Federrath2021} version of the \textsc{flash} code \citep{Fryxell2000,Dubey2008}, utilizing a second-order conservative MUSCL-Hancock 5-wave approximate Riemann scheme \citep{Bouchut2010,Waagan2011,Federrath2021} to solve the dimensionless 3D, ideal, isothermal, compressible MHD equations with a stochastic acceleration field acting to drive the turbulence with finite temporal correlation,
    \begin{align}
        \frac{\partial \rho}{\partial t} + \nabla\cdot(\rho \mathbf{v}) &= 0 \label{eq:continuity}, \\
        \rho\frac{\partial\mathbf{v}}{\partial t}  - \nabla\cdot\left[ \frac{1}{4\pi}\mathbf{B}\otimes\mathbf{B} - \rho \mathbf{v}\otimes\mathbf{v} - \left(c_s^2 \rho + \frac{B^2}{8\pi}\right)\mathbb{I}\right] &= \rho \mathbf{f},\label{eq:momentum} \\
        \frac{\partial \mathbf{B}}{\partial t} - \nabla \times (\mathbf{v} \times \mathbf{B}) &= 0,\label{eq:induction}\\
        \nabla \cdot \mathbf{B} &= 0, \label{eq:div0}
    \end{align}
    where $\mathbf{v}$ is the fluid velocity, $\rho$ is the gas density, $\mathbf{B} = B_0 \mathbf{\hat{z}} + \delta\mathbf{B}(t)$ is the magnetic field, with large-scale field $B_0 \mathbf{\hat{z}}$ and turbulent field $\delta\mathbf{B}$, $c_s$ is the sound speed, and $\mathbf{f}$ the stochastic turbulent acceleration source term that drives the turbulence. We solve the equations on a periodic domain of dimension $L^3\equiv \V$, discretised with between $16^3-288^3$ cells for the purpose of ensuring the convergence of numerical quantities used in this study. All of the quantities reported in the main text are for 53 unique simulations at $288^3$, which we list in \autoref{tb:simtab}. We show sample slices through some of the simulations in \autoref{fig:12_panel_gas_vars}. At a resolution of $288^3$ grid cells, the simulations have numerically converged $2^{\rm nd}$-moment statistics (see \autoref{sec:num_converge} for a numerical convergence test of the main statistics in this study), because the variance of the 3D turbulent fields is dominated by the lowest $k$-modes in the turbulence \citep[see also][]{Kitsionas2009_numerical_method_compare}.
    
    \subsection{Turbulent driving and sonic Mach numbers}
    The forcing term $\mathbf{f}$ follows an Ornstein-Uhlenbeck process that satisfies the stochastic differential equation,
    \begin{align} \label{eq:OU}
        \d{\hat{\vecB{f}}}(\vecB{k},t) = f_0(\vecB{k}) \mathbf{\mathbb{P}}(\vecB{k}) \d{\vecB{W}}(t) - \hat{\vecB{f}}(\vecB{k},t)\frac{\d{t}}{\tau},
    \end{align}
    where $\hat{\vecB{f}}(\vecB{k},t)$ is the Fourier transform of $\mathbf{f}$, with correlation time, such that $\hat{\vecB{f}}(\vecB{k},t) \sim f_0(\vecB{k}) \exp\left\{ - t/ \tau \right\}$ and $\tau = \ell_0/\Exp{v^2}_{\V}^{1/2} = L/(2 c_s \M)$ where $\ell_0 = L/2$ is the energy injection scale. Every $\tau$ the driving field loses $1/e$ of its previous structure. By controlling $\tau$ we are able to set $0.5 \lesssim \M \lesssim 10$, encapsulating the $\M$ values of supersonic molecular gas clouds in the interstellar medium \citep[e.g.,][]{Schneider2013,Federrath2016_brick,Orkisz2017,Beattie2019b} as well as in the sub-sonic, diffuse medium \citep[e.g. ][]{Kritsuk2017,Marchal2021_wnm_turbulence}. $\d{\vecB{W}}(t)$ is a Wiener process which draws delta-correlated random Gaussian increments from $\mathcal{N}(0,\d{t})$, a mean-zero Gaussian distribution with variance $\d{t}$, which is then projected onto $\vecB{f}$ isotropically in $k$-space with amplitude $f_0(\vecB{k})$. A filter is chosen such that the driving spectrum is concentrated at $|\mathbf{k}L/2\pi|=2$ and falls off to zero with a parabolic spectrum between $1 \leq |\mathbf{k}L/2\pi| \leq 3$. The projection is performed using the projection tensor
    \begin{align}
    \mathbb{P}_{ij} = \overbrace{\eta \left(\delta_{ij} + \frac{k_ik_j}{|k|^2} \right)}^{\rm{solenoidal\; modes}} + \underbrace{(1 -\eta)\frac{k_ik_j}{|k|^2}}_{\rm{compressive \; modes}}
    \end{align}
    where $\delta_{ij}$ is the Kronecker delta tensor. We control the contribution from each of the driving modes, indicated with the annotations for the two terms in the projection tensor, through the $\eta$ parameter. For $\eta = 1$ we obtain purely solenoidal driving $(\nabla \cdot \mathbf{f}=0)$, and $\eta = 0$ produces purely compressive driving  $(\nabla \times \mathbf{f}=0)$ \citep[see][for a detailed discussion of the driving]{Federrath2008,Federrath2009,Federrath2010,Federrath2022_turbulence_driving_module}. We choose to inject an equal amount of energy in both compressive and solenoidal modes, a ``natural mix", by setting $\eta = 0.5$ \citep[see][for turbulence driving details]{Federrath2008,Federrath2009,Federrath2010}. A natural mixture of modes is most appropriate for simulating ISM turbulence because driving mechanisms\footnote{Note that we are referring to \textit{driving} and not \textit{momentum}, i.e., gravitational collapse is a compressive source, but vorticity and compressible modes both are generated in the momentum field \citep{Higashi2021_turbulent_modes_in_collapse}.} are diverse, for example, supernova shocks (compressive), internal instabilities in the gas (solenoidal), gravity (compressive), galactic shear (solenoidal), ambient pressure from the galactic environment (compressive) or stellar feedback (compressive or solenoidal) \citep{Brunt2009,Elmegreen2009IAUS,Federrath2015_inefficient_SFR,Krumholz2016,Grisdale2017,Jin2017,Kortgen2017,Federrath2017IAUS,Colling2018,Schruba2019,Lu2020}\footnote{See Figure~2 in \citet{Sharda2021_driving_mode} for a catalog of sources that have been classified by different values of ``turbulent driving parameter", which is, in principle, an indirect measurement of $\eta$ (see \citealt{Federrath2009} for an empirical fit that maps one to the other). What is clear is that different turbulence driving mechanisms excite different modes, which may also depend on the galactic environment and/or time.}. 
    
    \subsection{Initial conditions, magnetic field and critical balance}
    
    $\Mao$ is set by fixing $B_0$, which we do when we set up the turbulent boxes, and using the definition of the mean-field Alfv\'en velocity, $\vao = B_0 / (4\pi\rho_0)^{1/2}$ with respect to the mean field $B_0$, and $\M$. Thus, $\Mao = c_s \M / \vao = 2c_s\sqrt{\pi\rho_0}\M/B_0$. We vary this value for each of the simulations between $10^{-2} \lesssim \Mao \lesssim 10^{2}$. $\Mao$ is almost a constant in time, since $B_0$ is constant, but because $\M$ fluctuates, so does $\Mao$. Regardless, we find that the desired value of $\Mao$ and the measured value closely match, as shown in the 2$^{\rm nd}$ column of \autoref{tb:simtab}. The initial velocity field is set to $|\vecB{v}(x,y,z,t=0)|=0$, with units $c_s=1$, the density field $\rho(x,y,z,t=0)/\rho_0=1$, with units $\rho_0=1$ and $|\delta\vecB{B}(x,y,z,t=0)|/(c_s\rho_0^{1/2}) = 0$, with units $c_s\rho_0^{1/2} = 1$. The turbulent magnetic field evolves self-consistently with the MHD fluid equations and satisfies $\Exp{\delta\vecB{B}(t)}_{\V}=0$, or more generally, $\Exp{\vecB{B}(t)}_{\V}=B_0$. This means $\partial_x B_0 = \partial_y B_0 = \partial_z B_0 = \partial_t B_0 = 0$, by construction. To ensure that the magnetic field is divergence-free, we use the parabolic $\nabla\cdot\vecB{B}$ diffusion method described by \citet{Marder1987_fluxcleaning}.
     
    Classifying the turbulence as weak or strong is beneficial for understanding the underlying turbulence phenomenology and statistics \citep{Sridhar1994_weak_turbulence,Goldreich1995,Perez_2008_weak_and_strong_turb,Boldyrev2009_spectrum_of_weak_turb,Oughton2020_review,Schekochihin2020_bias_review}. The critical balance parameter on the driving scale is $\zeta_{\rm crit} \sim t_{\rm alfven}/\tau \sim (\ell_{\parallel}/v_{A0}) (\ell_0 / c_s \M)^{-1}$ \citep{Goldreich1995,Oughton2020_review}, where $\ell_{\parallel} = \ell_0$ for isotropic driving, and hence $\zeta_{\rm crit} \sim c_s \M / \vao = \Mao$. Based on these simple arguments, on the outer scale of the turbulence, the $\Mao < 1$ ensemble of simulations may be considered weak (wave turbulence), and the $\Mao \geq 1$ may be considered strong. However, we urge caution in interpreting our simulations under the strong and weak phenomenologies in relation to the parameters chosen for our simulations, because even though the $\Mao < 1$ simulations may seem weak, we are injecting compressible modes isotropically in $k$-space into the turbulence to mimic astrophysical sources of driving. This creates strong shocks and rarefactions that form along field lines in the $\Mao < 1$ turbulence, which are discussed more in \autoref{sec:va_fluctuations} and visualized in \autoref{fig:12_panel_gas_vars_2}. The presence of shocks (which are non-local in $k$-space) that evolve on very short timescales \citep[e.g, shorter than the local sound crossing time,][]{Robertson2018,Mocz2018} means that in many regions of the plasma non-linear effects cannot be neglected. This may prevent the turbulence from ever becoming truly weak.
    
    \subsection{Stationarity and collecting statistics}
    
    We run the simulations for $10\tau$, and report statistics from quantiles for the $50^{\rm th}$, $16^{\rm th}$ and $84^{\rm th}$ percentiles of the time distributions over the last $5\tau$. This (1) ensures that the sub-Alfv\'enic, large-scale field simulations are statistically stationary, i.e., $\Exp{\vecB{X}(t)}_{\V} = \Exp{\vecB{X}(t+\Delta t)}_{\V}$, for an arbitrary field variable $\vecB{X}$, and increment in time $\Delta t$, which takes roughly $5\tau$ for those experiments \citep{Beattie2021_spdf}\footnote{We note that super-Alfv\'enic turbulence takes roughly $2\tau$ to approach a statistically stationary state, similar to purely hydrodynamic turbulence \citep{Price2010}.} and (2) makes all of our results robust to temporally intermittent events, which are ubiquitous in turbulence experiments. Unless explicitly written, we will average every statistic in this study, including 1D and 2D PDFs in \autoref{sec:dens}-\autoref{sec:correlations}, to make sure our results are as robust as possible. Throughout the study we use a naming convention for our simulations whereby the value following the \texttt{M} gives the target $\M$ (with omitted decimal points) and the value following \texttt{MA} gives the target $\Mao$ -- thus run \texttt{M10MA01} is one where we set the large-scale magnetic field Alfv\'en Mach number to $\Mao=0.1$ and tune the correlation time of the stochastic forcing to produce a sonic Mach number of $\M=10$. Now we turn to the results of this study.

    \section{Ionic Alfv\'en velocity fluctuations}\label{sec:va_fluctuations}
    Consider the dimensionless magnitude of the ionic Alfv\'en velocities in a magnetized, compressible plasma,
    \begin{align}
        \vai/c_s &= \frac{B/(c_s \rho_0^{1/2})}{\sqrt{4\pi\chi(\rho/\rho_0)}}, 
    \end{align}
    The correlation between the gas density and the ionization  fraction is, 
    \begin{align}
        \chi = \chi_0 (\rho/\rho_0)^{-1/p},
    \end{align}
    where $\chi_0$ is the mass ionization fraction at density $\rho_0$. When $p\rightarrow \infty \implies \chi = \chi_0$, the equilibrium time between ionizing the plasma $t_{\rm ion}$ and a typical density fluctuation $t_{\rho/\rho_0}$ is comparable $t_{\rm ion}\sim t_{\rho/\rho_0}$ (warm atomic gas), and for $p=2$, $t_{\rm ion} \ll t_{\rho/\rho_0}$ (cold molecular or atomic gas), following \autoref{eq:ion_equilibrium}; ideal MHD is $\chi_0 = 1, p\to\infty$. For a general $p > 0$,
    \begin{align}
        \vai/c_s &= \frac{B/(c_s \rho_0^{1/2})}{\sqrt{4\pi\chi_0(\rho/\rho_0)^{(p-1)/p}}} = \frac{e^{\lB}B_0/(c_s \rho_0^{1/2})}{\sqrt{4\pi\chi_0(\rho/\rho_0)^{(p-1)/p}}}, 
    \end{align}
    where 
    \begin{align}
        B = e^{\lB }B_0, \;\text{and}\;\lB = \ln(B/B_0).
    \end{align}
    We make this change of variables to bring out the symmetry in the log transform of $\vai$,
    \begin{align}\label{eq:va_joint}
        \lvai  &= \lB - \frac{p-1}{2p}s,
    \end{align}
    where $s = \ln(\rho/\rho_0)$, $\lvai = \ln(v_{A,\rm{ion}}/v_{A0,\rm{ion}})$ and $v_{A0,\rm{ion}} = B_0/(c_s \sqrt{4\pi\rho_0\chi_0})$. Now $\lvai$ is simply the addition of two random variables, $\lB$ and $-[(p-1)/(2p)]s$. This means the $\lvai$ variance is 
    \begin{align} \label{eq:alfven_var}
        \sigma_{\lvai}^2 &= \sigma_{\lB}^2 + \left(\frac{p-1}{2p}\right)^2\sigma_s^2 - \frac{p-1}{p}\sigma_{\lB, s},
    \end{align}  
    where $\sigma_{\lB, s}$ is the spatial covariance between $\lB$ and $s$,
    \begin{align}
        \sigma_{\lB, s} &= \big<\big(\lB - \big<\lB\big>_{\V}\big)\big(s - \big<s\big>_{\V}\big)\big>_{\V}.
    \end{align}
    Thus, if we want to model the PDF and variance of $\lvai$, and in turn, $\vai$, we must understand $\sigma_{\lB}^2$, $\sigma_s^2$, and the covariance between them. Hence, before we focus our attention on Alfv\'en speed fluctuations themselves, we first consider the three necessary ingredients: density and magnetic field fluctuations and the correlations between them. We start with density fluctuations.
    
    \subsection{Density fluctuations}\label{sec:dens}
    
    \begin{figure}
        \centering
        \includegraphics[width=\linewidth]{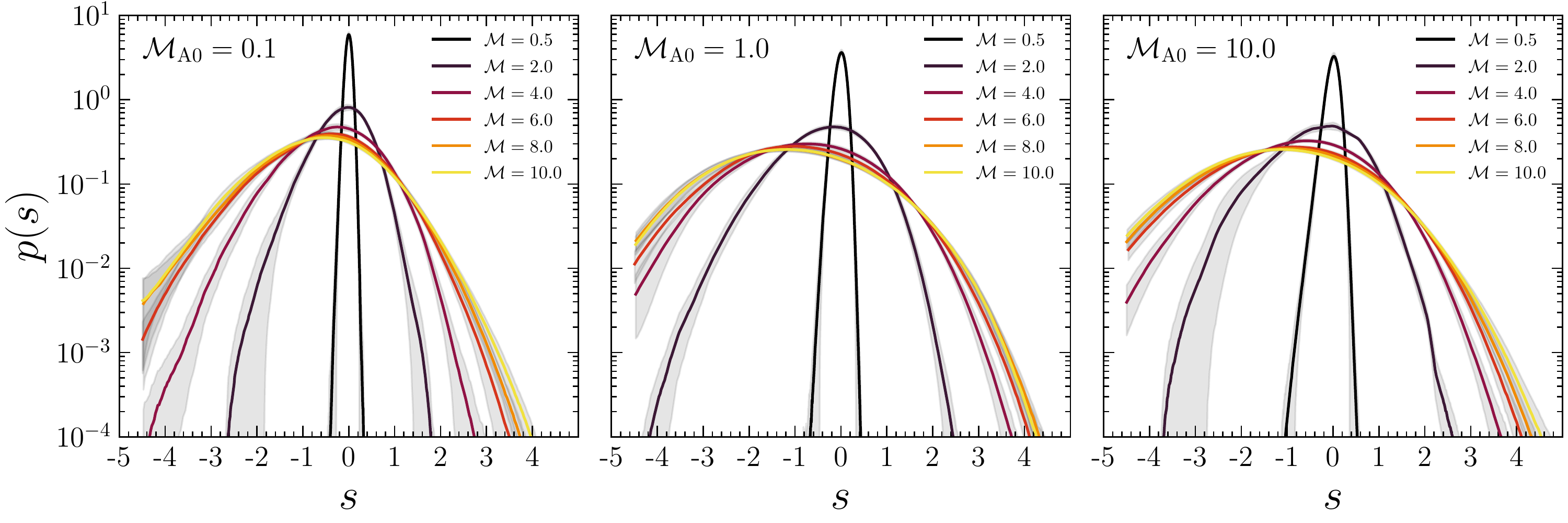}
        \caption{The logarithmic density PDFs for the sub-Alfv\'enic, $\Mao =0.1$, simulations (left), trans-Alfv\'enic, $\Mao =1.0$ simulations (middle), and super-Alfv\'enic, $\Mao = 10.0$, simulations (right), colored by $\M$. Between $0.5 \leq \M \leq 4$ we find monotonic growth in the spread of the PDF, showing that as the flow becomes more compressible the density fluctuations increase in magnitude and dispersion. Beyond $\M \approx 4-6$, complex correlations between the density and magnetic field suppress the largest over- and under-densities \citep{Molina2012_dens_var, Beattie2021_multishock,Beattie2021_spdf}, and the variance of the PDF stops growing.}
        \label{fig:s_pdfs}
    \end{figure}    
    
    \begin{figure}
        \centering
        \includegraphics[width=\linewidth]{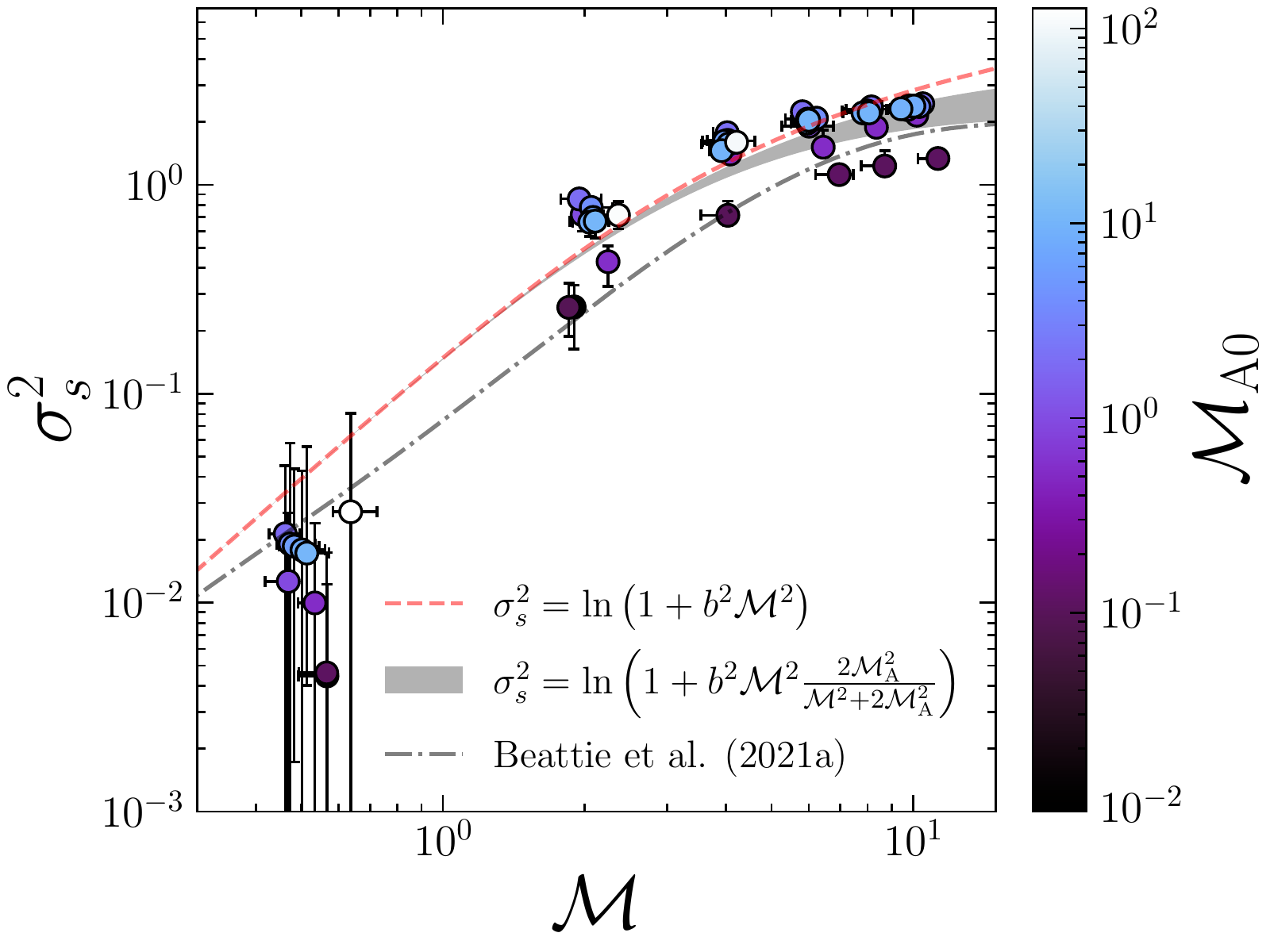}
        \caption{The logarithmic density variance, $\sigma_s^2$ as a function of $\M$, colored by $\Mao$. We show the hydrodynamical model of the variance, $\sigma_s^2 = \ln(1 + b^2\M^2)$, \citep{Federrath2008,Federrath2010} shown with the red, dashed line, the \citet{Molina2012_dens_var} model, $\sigma_s^2 = \ln(1 + b^2\M^2[2\Ma/(\M^2 + 2\mathcal{M}_{\text{A}}^2)])$, which corrects for magnetic pressure, evaluated between $\Mao = 6-10$ with the gray band, and the \citet{Beattie2021_multishock} model, which corrects for the anisotropic nature of sub-Alfv\'enic density fluctuations, evaluated at $\Ma = 0.1$ with the gray dashed-dotted line. The variance models describe the data well for $\M \gtrsim 2$ but become worse as $\M$ decreases, where the compressible modes are unable to create significant density fluctuations.}
        \label{fig:s_var}
    \end{figure}    
    
    In \autoref{fig:s_pdfs} we show the $s$-PDFs from the $\Mao=0.1$ (left), $\Mao=1.0$ (middle) and $\Mao=10.0$ (right) simulations, colored by $\M$. From $\M = 0.5 - 4$, the $s$-PDF monotonically increases with $\M$, showing how the amplitude and dispersion in density fluctuations becomes greater as the compressibility in the magnetized plasma increases, regardless of $\Mao$, as shown previously in \citet{Molina2012_dens_var} and \citet{Beattie2021_multishock,Beattie2021_spdf}. At $\M \gtrsim 4$ the spread of the $s$-PDFs begins to slow, which is not found in hydrodynamical compressible turbulence \citep[see Figure~1 in ][]{Price2011} but has been explained in \citet{Molina2012_dens_var} and \citet{Beattie2021_spdf} as due to correlations between the magnetic field and density field growing, and ``cushioning" the plasma, preventing it from creating strong over-and-under-densities. These correlations become important for $\va$, which we will discuss and explore in more detail in \autoref{sec:va_fluctuations}. Overall, as $\M$ increases, the PDFs become more negatively skewed as the plasma becomes rich in volume-filling rarefactions. This is because the characteristic thickness of an over-dense region scales with $\sim \ell_0/\M^2$ \citep{Padoan1997_imf,Molina2012_dens_var,Robertson2018,Mocz2018}, and hence, as $\M$ increases the shocked gas becomes less volume-filling (skewing the PDFs), but contains most of the mass in the plasma \citep{Robertson2018,Beattie2021_spdf}.
    
    In \autoref{fig:s_var} we show $\sigma_s^2$ for the entire ensemble of simulations utilised in this study, plotted as a function of $\M$. For reference, we also plot the relation from hydrodynamical supersonic turbulence theory (dashed, red), 
    \begin{align}
        \var{s} = \ln\left(1 + b^2\M^2\right),
    \end{align}
    \citep{Federrath2008,Federrath2010}, super-Alfv\'enic MHD turbulence theory (gray band), 
    \begin{align}
        \var{s} = \ln\left(1 + b^2\M^2\frac{2\Ma}{\M^2 + 2\mathcal{M}_{\text{A}}^2}\right),
    \end{align}
    assuming $\rho \propto B^{1/2}$ \citep{Molina2012_dens_var} evaluated between $\Mao = 2 - 10$, and sub-Alfv\'enic MHD turbulence theory (dot-dashed, gray),
    \begin{align}
        \var{\rho/\rho_0} &=  \overbrace{\frac{\V_0 b^2 \M^2}{\sqrt{1 + \left(\frac{\M}{\M_{\rm c}}\right)^4}}}^{\text{variance along $\Bo$}} + \underbrace{\frac{\sqrt{1 + \left(\frac{\M}{\M_{\rm c}}\right)^4}-f_0}{2 \sqrt{1 + \left(\frac{\M}{\M_{\rm c}}\right)^4}}\left[ \sqrt{\left(1 + 2\frac{\MaO{2}}{b^2\M^2}\right)^2 + 8\MaO{2}} -\left(1 + 2\frac{\MaO{2}}{b^2\M^2}\right) \right]}_{\text{variance across $\Bo$}},\\ \label{eq:2shockModel}
        \var{s} &= \ln\left(1 + \var{\rho/\rho_0}\right),
    \end{align}
    where, $\V_0$ is the volume-filling fraction of shocked gas along $\Bo$ in the trans-sonic limit and $\M_{\rm c}$ is the critical $\M$ where the volume-filling fraction of the gas along $\Bo$ becomes negligible \citep{Beattie2021_multishock}. As with \autoref{fig:s_pdfs}, regardless of $\Mao$, $\sigma_s^2$ shows qualitatively similar behavior -- becoming monotonically more weakly dependent upon $\M$ as $\M$ increases, following the logarithmic trend captured by the density variance models. However, as $\M$ increases, the sub-Alfv\'enic experiments settle to values roughly a factor of $2$ lower in $\sigma_s^2$ than both the super-Alfv\'enic experiments and the prediction from the hydrodynamical variance model (shown with the red-dashed lines). For these sub-Alfv\'enic plasmas, \citet{Beattie2021_multishock} attributed the stronger flattening of the variance to fluctuations along the $\Bo$\footnote{Note that we have naturally introduced an orientation for the density fluctuations. This is probably the most significant effect that the magnetic field has on the density variance. The overall magnitude does not change significantly, but through flux-freezing, the magnetic field acts as a scaffold for the fluctuations, making them highly anisotropic along and across $\Bo$ \citep[see Figure~2 in][]{Beattie2020}.} (uncorrelated, $\rho \propto B^0$) becoming extremely low volume-filling, and the fluctuations across the field lines ($\rho \propto B$) becoming the dominant source of volume-weighted variance. Since $\rho \propto B$ fluctuations have to perturb the magnetic field, and because $\vecB{B}_0$ is very strong in the sub-Alfv\'enic regime, these are weak fluctuations that do not grow the variance as fast as the $\rho \propto B^0$ fluctuations.

    The key point is that, regardless of $\M$ and $\Mao$, the lognormal model for the density PDF is a reasonable (and practical) approximation to make. Equipped with this knowledge, we now move on to the second ingredient we need to understand the nature of Alfv\'en wave fluctuations in compressible MHD turbulence — the magnetic field fluctuations.   
    
    \begin{figure}
        \centering
        \includegraphics[width=\linewidth]{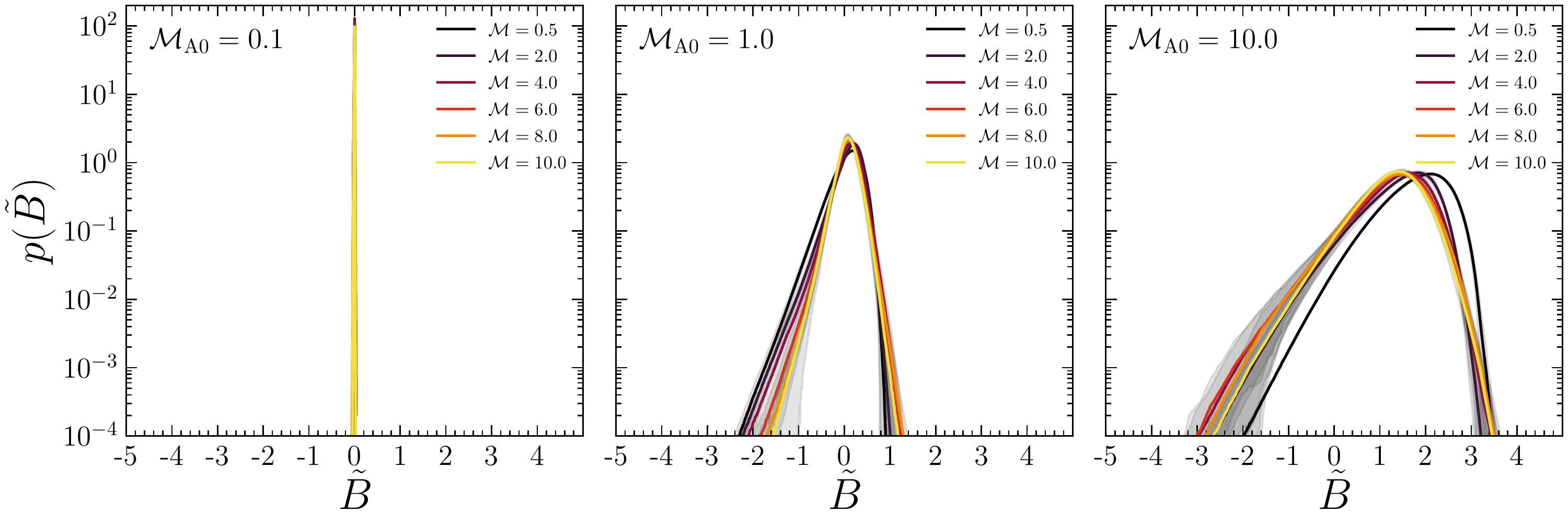}
        \caption{The same as \autoref{fig:s_pdfs}, but for the logarithmic magnetic field magnitudes. For $\Mao < 1$, the distribution is dominated by $\Bo$, with negligible fluctuations. As $\Mao$ increases, and $B_0$ decreases, the turbulent component of the field grows and the distribution spreads out and becomes negatively skewed. The tails and median of the distribution are weakly dependent upon $\M$. The negative skewness grows with $\Mao$ (weakening $B_0$), indicating that more of the simulation volume is filling with highly-tangled magnetic fields, which produce magnetic field magnitudes larger than $B_0$.}
        \label{fig:b_pdf}
    \end{figure}    
    
    \subsection{Magnetic field fluctuations}\label{sec:magnetic}
    
    In a similar treatment as the last section, we plot a representative sample of the logarithmic magnetic magnitude PDFs in \autoref{fig:b_pdf} on the same scale as the PDFs shown in \autoref{fig:s_pdfs}. Unlike the $s$-PDFs, the $\lB$-PDFs show a very strong dependence upon the strength of the large-scale magnetic field (or likewise, $\Mao$) and only a weak dependence upon $\M$. In the sub-Alfv\'enic regime (left panel) the magnetic field resembles a delta function centered on the value of large-scale field, $\ln(B/B_0)=0$. As $\Mao$ increases, and the strength of the large-scale field decreases the turbulent component of the magnetic field is able to grow, and we can observe some structure in the $\lB$-PDFs. At $\Mao = 1$, $\Bo$ is still dominant ($\delta B^2 / B_0^2 = 1$ at $\Mao = 2$, \citealt{Beattie2020_mag_fluc,Beattie2022_inprep_energybalance}), however small asymmetries between volumes with $B > B_0$ and $B < B_0$ develop and negatively skew the PDF. These features become extremely pronounced at $\Mao = 10.0$ \citep[similar non-Gaussian features are very pronounced for $\Mao = \infty$; ][]{seta2021saturation}, where $\delta B^2 \gg B_0^2$, and hence, when the nonlinear terms in the fluid equations are the strongest \citep{oughton_priest_matthaeus_1994}. 
    
    Not only are the higher-order moments of the PDFs growing with $\Mao$, but the volume-weighted $\lB$-PDFs shift to higher values of $\lB$, with most values $B > B_0$, as $\Mao$ increases. This means that when the large-scale field is weak, the turbulence is able to feed the fluctuating magnetic field with energy, growing most of the field beyond $B_0$. As shown in Figure~2 of \citet{Beattie2020}, the topology of the magnetic field becomes extremely tangled and complex. It is well known that tangling, or knotting the field, by increasing the number of crossings between flux tubes grows the magnetic energy linearly with number of crossings\footnote{Strictly speaking, the lower bound of the magnetic energy, $E_{\rm min}$ and the minimum number of crossings for any topologically equivalent magnetic fields, $C_{\rm min}$, respectively, such that $E_{\rm min}\geq C_{\rm min}[16/(\pi \V)]^{1/3}$.} \citep{Freedman1991_magnetic_field_topology_crossings,Seligman2022_twisted_b_fields}. The net result of this is that the magnetic field becomes tangled, paths along field lines become volume-filling, and $B > B_0$ in most of the fluid volume. The middle column of \autoref{fig:12_panel_gas_vars} shows exactly this: as $\Mao$ increases from 0.1 to 10.0, the amount of pink $(B/B_0 > 1)$ compared to blue $(B/B_0< 1)$ increases and becomes volume-filling. In contrast to the $s$ variance, there is much less theory on the magnetic field variance. However, recent works have had some success in understanding the $B/B_0$ (linear magnetic field) variance. Next we summarise those results and compare them to our simulation data.
    
    \begin{figure}
        \centering
        \includegraphics[width=\linewidth]{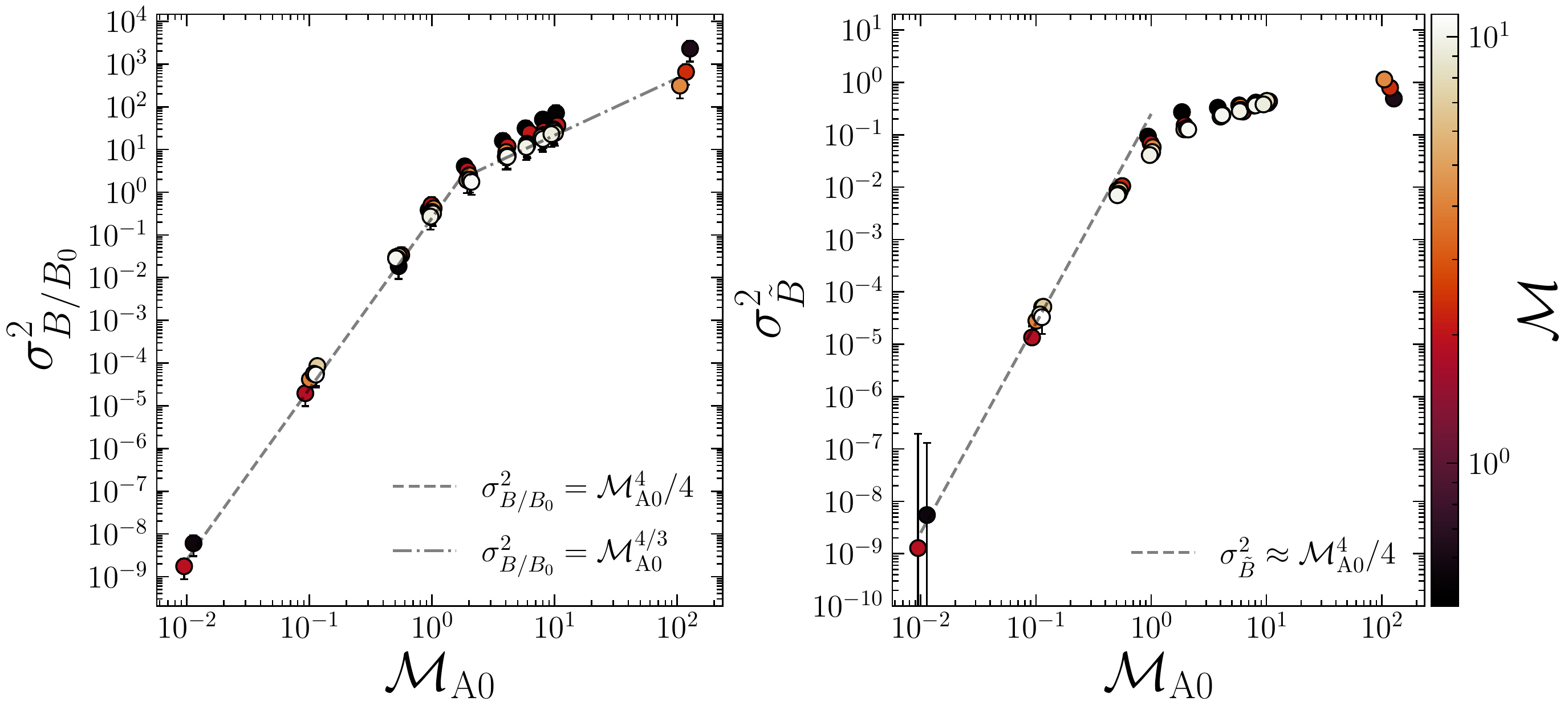}
        \caption{The variance of $B/B_0$ (left) and $\ln(B/B_0)$ (right) as a function of $\Mao$, colored by $\M$. In the left panel, we plot the $B/B_0$ variance models (gray-dashed lines) derived through energy balance arguments in \citet{Beattie2022_inprep_energybalance}, which capture both the sub-Alfv\'enic and super-Alfv\'enic regime. For $\sigma^2_{\lB}$, in the right panel, we use the delta method to derive the variance in the sub-Alfv\'enic regime, which, to linear order, is equal to $\sigma^2_{B/B_0}$.}
        \label{fig:b_variance}
    \end{figure}        
    
    We plot the variance of both the linear, large-scale field normalized magnetic field $B/B_0$ (left panel) and logarithmic magnetic field (right panel) of \autoref{fig:b_variance}. Both plots show a systematic power law increase with $\Mao$, with a break scale between $\Mao \approx 1-2$. This marks the energy equipartition transition between the turbulent magnetic field energy and large-scale magnetic field energy. For the linear field, we utilize the models developed in \citet{Beattie2022_inprep_energybalance}. They find that by taking the $2^{\rm nd}$ moments of the energy balance between the magnetic and kinetic energy,
    \begin{align} \label{eq:energy_balance_rms}
        \overbrace{\frac{1}{8\pi c_s^2 \rho_0}\Exp{\left(2\dB\cdot\Bo + \delta B^2\right)^2}_{\V}^{1/2}}^{\text{2$^{\rm nd}$ moments of the turbulent magnetic energy}} \propto \underbrace{\frac{1}{2}\Exp{\left(\frac{\delta v}{c_s}\right)^4}^{1/2}_{\V}}_{\text{2$^{\rm nd}$ moments of the turbulent kinetic energy}}.
    \end{align} 
    relations between the rms magnetic field, including the effects of the large-scale field via the $\dB\cdot\Bo$ term in the above equation, can be derived without any use of fitting parameters. For the variance of $B/B_0$, this results in,
    \begin{align}
        \sigma_{B/B_0}^2 &= \begin{cases}
        \mathcal{M}_{A0}^4/4 & \text{ if } \Mao \leq 2, \label{eq:mag_var}\\
        \mathcal{M}_{A0}^{4/3} & \text{ if } \Mao > 2, 
        \end{cases}
    \end{align}
    which show excellent agreement with the simulation data, across all $\Mao$ and $\M$. Note also that $\Ma \approx \Mao$ when $\Mao \leq 2$ \citep{Beattie2022_inprep_energybalance}, and $\Ma$ is an observable quantity using \citet{Davis1951}, \citet{Chandrasekhar_fermi_1953} and \citet{Skalidis2020} starlight polarization methods, hence, in principle, $\sigma_{B/B_0}^2$ is a derivable quantity from observations. We discuss this more in \autoref{sec:discussion}. However, \autoref{eq:va_joint} critically depends upon the $\lB$ variance, and not the $\sigma_{B/B_0}^2$. Using the delta method \citep[see, e.g., ][ for a summary of the method]{Hoef2012_deltamethod}, one can approximate the moments of functions of random variables utilizing the Taylor expansion of the function. To linear order, the variance is, $\text{Var}\left\{ f(X) \right\} = \left[\partial_X f(\Exp{X})\right]^2\text{Var}\left\{ X \right\}$,where $\text{Var}\left\{ X\right\}$ is the variance operator applied to random variable $X$. In our case, $X= B/B_0$ and hence $\Exp{B/B_0} = 1$, and $\partial_{B/B_0} \ln(\Exp{B/B_0}) = 1$. This means, to linear order, $\sigma_{\lB}^2 = \sigma_{B/B_0}^2$. But, the delta method only works well for $\text{Var}\left\{ B/B_0 \right\}/\Exp{B/B_0} < 1$, hence we are only able to apply this approximation in the sub-Alfv\'enic regime, where this condition is met. We plot this function in the right panel of \autoref{fig:b_variance}, which shows reasonable agreement with the data up until $\sigma_{\lB}^2\approx 0.1$, which corresponds to $\Mao = 1$.

    \subsection{Covariance between the magnetic and density fluctuations}\label{sec:correlations}
    
\begin{figure*}
        \centering
        \includegraphics[width=\linewidth]{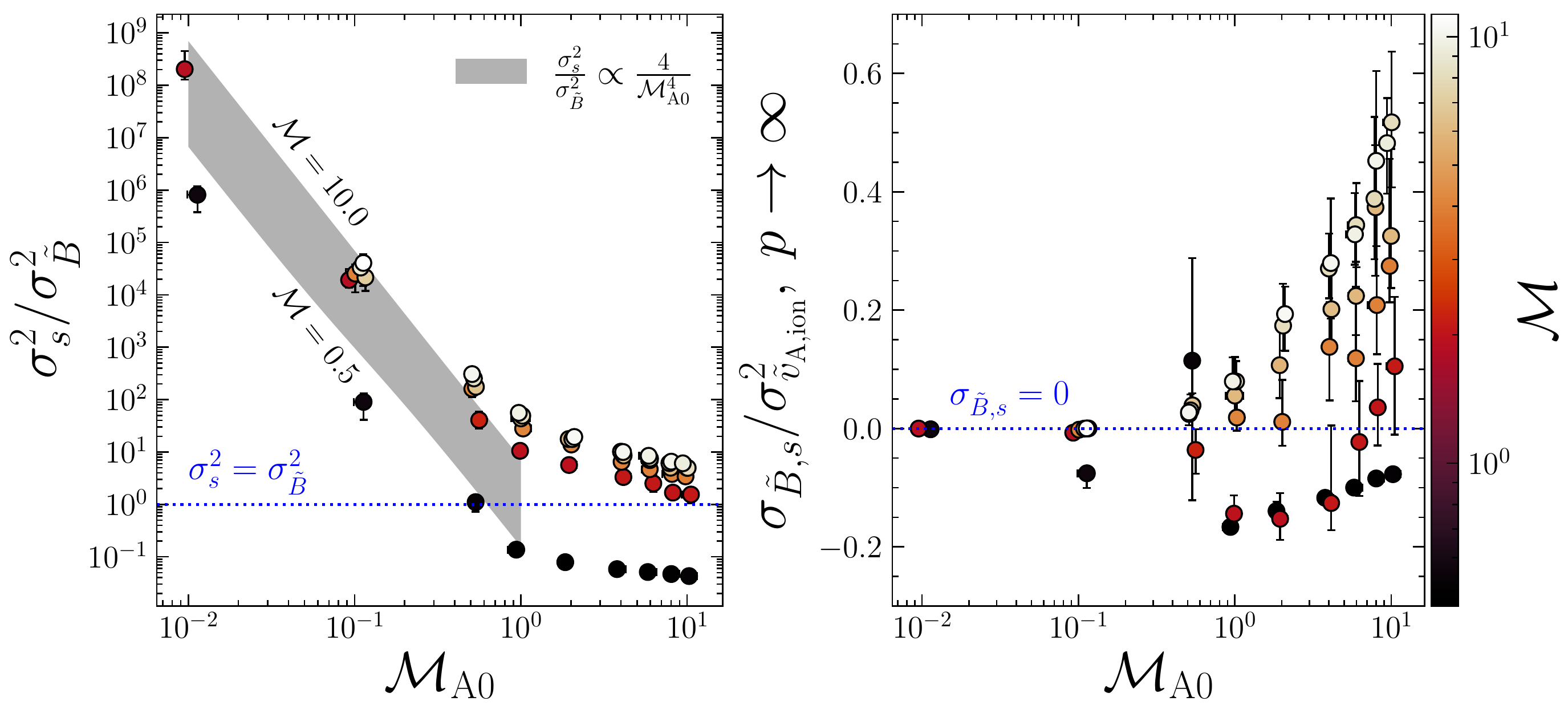}
        \caption{\textbf{Left:} The logarithmic density and magnetic field variance ratio as a function of $\Mao$, colored by $\M$. In $\Mao < 1$ regime, the $s$ fluctuations are many orders of magnitude larger than the magnetic field fluctuations. As $\Mao$ increases, the $\lB$ field fluctuations contribute more strongly to the ratio and the ratio flattens. If $\M < 1$, then at high-$\Mao$ the $\lB$ fluctuations dominate, and vice versa for $\M > 1$. Using the gray band, we plot the ratio between \autoref{eq:2shockModel} and \autoref{eq:mag_var}, which shows $\var{s}/\var{\lB}\propto \MaO{-4}$, for $\Mao <1$. \textbf{Right:} The same as the left panel but for the covariance of the logarithmic magnetic field $\lB$ and density $s$ weighted by the variance of the logarithmic ion Alfv\'en velocities (in the ideal limit $p\rightarrow\infty$ and $\chi = \chi_0 = 1$). This statistic shows the magnitude of the correlations between the $\lB$ and $s$ fields for the different strengths in large-scale magnetic field. At low-$\Mao$, the correlations are negligible, and hence $\lB$ and $s$ are independent. However, for $\Mao \approx 1$, the correlations contribute significantly to $\sigma^2_{\lvai}$.}
        \label{fig:cov_Bs}
    \end{figure*}

    In the last two sections we developed an understanding of the global fluctuations (the variance) of the logarithmic density and magnetic field fluctuations. It may already be apparent that in the sub-Alfv\'enic regime the magnetic field fluctuations are completely negligible, and for $\lvai$, the density fluctuations are going to be dominant. To make this more quantitative, we plot the ratio of the $s$ and $\lB$ variances in the left panel of \autoref{fig:cov_Bs}. As expected $\sigma_s^2 \gg \sigma_{\lB}^2$ in the sub-Alfv\'enic regime, by up to nine orders of magnitude at $\Mao=0.01$ down to two orders of magnitude at $\Mao=1.0$. Between $\M=2$ and $\M=10$ there is roughly an order of magnitude difference, with the $s$ fluctuations playing a larger role with increasing $\M$. This highlights the importance of compressibility in the sub-Alfv\'enic regime, in particular, where the $\lB$ fluctuations have orders of magnitude less power than the $s$ fluctuations. 
    
    Using the \citet{Beattie2021_multishock} model for the $s$ variance, \autoref{eq:2shockModel}, which we plotted in \autoref{fig:s_var}, and the $\lB$ variance, \autoref{eq:mag_var}, we use from \citet{Beattie2022_inprep_energybalance}, for the sub-Alfv\'enic regime, we get,
    \begin{align}
        \frac{\sigma_s^2}{\sigma_{\lB}^2} \propto \frac{4}{\MaO{4}}, \; \text{ if } \Mao \leq 1,
    \end{align}
    where the proportionality factor is \autoref{eq:2shockModel}, and for when the delta method for approximating $\var{\lB}$ is valid. We plot this model evaluated at $\M = 10$ and $\M = 0.5$ with the gray band in \autoref{fig:cov_Bs}. The slope $\sim \MaO{-4}$ looks accurate, but the exact offset, which is a function of the $\sigma_s^2$ prescription, does not capture the subsonic regime properly, which makes sense because the $\sigma_s^2$ models are valid for the $\M > 1$ regime, as discussed in \autoref{sec:dens}. Regardless, the model describes the general trend, and is useful in that we now know $\var{s}/\var{\lB} \propto \MaO{-4}$, which highlights that in the sub-Alfv\'enic regime, it is the density fluctuations that are leading order in \autoref{eq:va_joint}. 
    
    \begin{figure*}
        \centering
        \includegraphics[width=\linewidth]{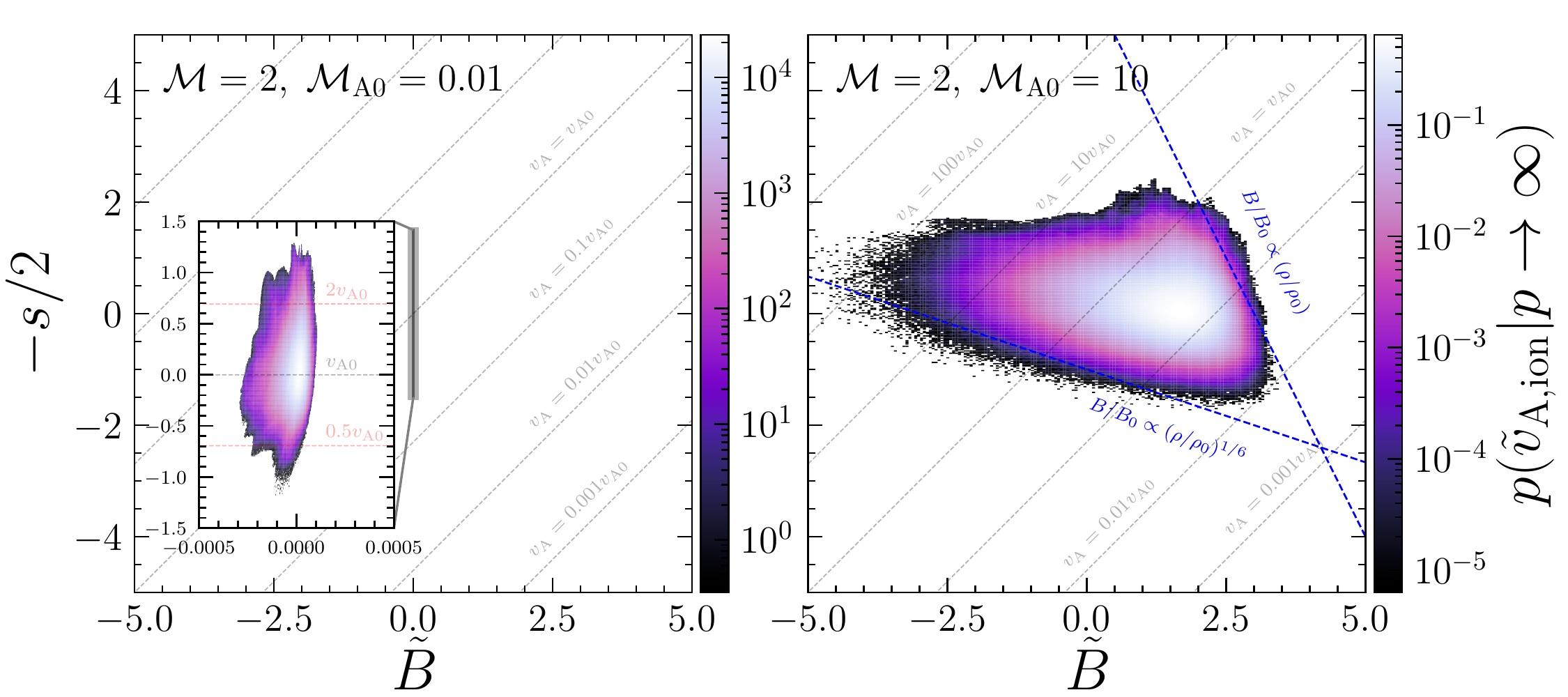}
        \caption{The joint $\lB \equiv \ln(B/B_0)$ and $-s/2 \equiv -1/2 \ln(\rho/\rho_0)$ PDFs, which we demonstrate is the $\lvai$-PDF in the $\chi=\chi_0=1$ and $p\rightarrow\infty$ limit, \autoref{eq:va_joint_limit}. We show gray lines of constant $\va$ in units of $\vao$. Each line represents a change in $\va$ by an order of magnitude in $\vao$. \textbf{Left:} the joint PDF for the \texttt{M2MA001} simulation, showing that in the sub-Alfv\'enic regime the variation in $\lvai$ is determined solely by the density fluctuations, and that the $\lB$ and $s$ fluctuations can be treated as independent. The mechanism for creating density fluctuations independent of the magnetic field are from channel flows along the field lines that compress the fluid into perpendicular filaments along $\Bo$ \citep{Beattie2020,Beattie2021_spdf}. \textbf{Right:} the same as the left panel but for the super-Alfv\'enic \texttt{M2MA10} simulation, revealing significant variation in both the $\lB$ and $s$ fluctuations, with some rough correlations enveloping the PDF that we add to guide the eye.}
        \label{fig:joint_b_s}
    \end{figure*}    
    
    Of course comparing the first two terms in \autoref{eq:va_joint} is important to determine which is leading order, but if the covariance between the fields is large, then this will still complicate the modeling of $\var{\lvai}$. Hence, in the right panel of \autoref{fig:cov_Bs} we plot the ratio between the covariance of $\lB$ and $s$ and $\var{\lvai}$. This gives us an idea of the magnitude of the covariance contribution to $\var{\lvai}$. For $\Mao < 1$, we find that the covariance term is negligible, $\sigma_{\lB,s}\approx 0$. As discussed more qualitatively in \citet{Molina2012_dens_var} and \citet{Beattie2021_spdf} (and previously in \autoref{sec:dens}), as $\Mao$ gets larger, the covariance (correlations between the magnetic and density field) begins to become sensitive to $\M$. For the supersonic simulations, the covariance is positive, and makes the largest contributions for the highest $\M$, $\sigma_{\lB,s}\approx (0.5-0.6)\var{\lva}$. This is qualitatively similar to previous results involving measuring the correlations between density and magnetic properties of supersonic turbulence \citep[e.g.,][]{Burkhart2009,Yoon2016_pressure_balance}. Most likely this comes about from flux-freezing and compressive motions that lead to the over-dense, highly-magnetized regions that we show in the last row of \autoref{fig:12_panel_gas_vars}. Because of the extra magnetic pressure, and the strong correlations, this also means that it is hard to create large compressions, which is why the variance of the $s$-PDF grows more slowly than in hydrodynamical turbulence, as discussed in \autoref{sec:dens}. 
    
    In contrast, the sub(to-trans)sonic, super-Alfv\'enic experiments give rise to a negative covariance. This could mean either the magnetic field is strong in under-densities, or the field is weak in over-densities, or both. If one carefully analyses the $s$-PDF and the $\lB$-PDF, then one can see that it is under-densities where the magnetic field becomes strong, and if it were over-densities giving rise to a weak magnetic field the $s$-PDF would need to be skewed in the opposite direction given the negative skewness in the $\M = 0.5$ $\lB$-PDF. Previous experiments have concluded that this is due to a thermal, $(1/\M^2)\nabla c_s \rho$, and magnetic pressure, $(1/\mathcal{M}_{\text{A}}^2)\nabla B^2$, balance, $\Ma \approx \M$, that allows the fluid to equilibriate on shorter timescales than the correlation times in subsonic turbulence \citep{Yoon2016_pressure_balance}. This amounts to regions of strong magnetization evacuating the gas density until the thermal pressure back-reacts and an equilibrium is reached. Of course, when $\M \gg 1$, the thermal pressure becomes less important and the short correlation timescales in the turbulence do not give the plasma a large enough correlated time interval to undergo such a process. Based on \autoref{fig:cov_Bs}, $\M \approx 2-4$ defines the critical $\M$ for when the equilibration time is shorter than the correlation time of the turbulence.
    
    To understand these correlations (or lack thereof) further, we plot the time-averaged joint distributions of $-[(p-1)/2p]s$ and $\lB$ in \autoref{fig:joint_b_s} in the limit where $p \rightarrow \infty$ and $\chi = \chi_0 = 1$, which gives, 
    \begin{align}\label{eq:va_joint_limit}
        \lim_{p \rightarrow \infty}\lvai  &= \lB - \frac{1}{2}s,
    \end{align}
    to compare directly with our simulation data. The benefit of making this joint PDF, as opposed to $B$ versus $\rho$, or $B^2$ versus $\rho$, is that lines in this space have constant Alfv\'en velocity magnitudes, as shown in \autoref{eq:va_joint}, and the probability density is then exactly the probability density of $\lvai$, which is the quantity of interest in our study. Because $-s/2$ is negative, positive correlations in the plot go down the $-s/2$ axis. We plot the joint PDF for a highly sub-Alfv\'enic simulation, \texttt{M2MA001} on the left, and a super-Alfv\'enic simulation, \texttt{M2MA10}, on the right, with constant Alfv\'en velocities in units of the mean-field Alfv\'en velocities shown with gray contours in both plots, and blue lines for correlations between $B/B_0$ and $\rho/\rho_0$ that define the envelopes of the PDF for the super-Alfv\'enic data.
    
    To be able to visualize the structure in the sub-Alfv\'enic data, we require a zoom-in to see the variation in $\lB$, which we show with the inset panel. The PDF varies over $\mathcal{O}(10^{-4})$ in $\lB$ compared to $\mathcal{O}(10^{-1})$ for $s$, consistent with all of the other results presented in this study. As we found with the covariance, there is no systematic change orientation in the $-s/2-\lB$ plane for the sub-Alfv\'enic data, which means that $s$ and $\lB$ are independent. The physical reason why this is the case in the sub-Alfv\'enic regime is simple. Because the large-scale field is so strong, and is unable to be bent by the turbulence (sub-Alfv\'enic large-scale turbulence naturally implies $B_0^2 \gg \delta v^2$), through flux-freezing, density fluctuations can only form from compressible velocity channels up and down magnetic field lines\footnote{Note that there are still weak compressions perpendicular to field lines, which, in the framework of linear MHD theory, are from compressible fast magnetosonic modes. These form striations of weakly compressed gas running parallel to the magnetic field \citep{Tritsis2018,Beattie2020,Beattie2020_mag_fluc}.}, where both $B$ and $\rho$ are uncorrelated from one another. These channels are very important for turbulence in the ISM that may be sub-Alfv\'enic, because through the converging channel flows the plasma is able to compress the gas, essentially with hydrodynamical shock-jump conditions, $\rho/\rho_0\sim \M^2$ \citep{Mocz2018,Beattie2021_spdf}, allowing for very dense filaments that form perpendicular to $\Bo$, which undoubtedly become the sites of star formation in sub-Alfv\'enic star-forming molecular clouds \citep{Padoan1999,Chen2014,Abe2020,Chen2020,Bonne2020}. Clearly, based on the joint PDF, this is the sole mechanism for forming strong over-densities in the highly sub-Alfv\'enic regime, and hence, the sole mechanism for creating dispersion in the Alfv\'en velocities.
    
    The super-Alfv\'enic data shows a wide spread in $\lB$ and $s$ values, with a $B/B_0\propto\rho/\rho_0$ correlation at large $B$ amplitudes, corresponding to compressions perpendicular to magnetic field lines \citep[e.g.,][]{Tritsis2015}, and a weak $B/B_0\propto(\rho/\rho_0)^{1/6}$ correlation at low-$s$, possibly due to the tangled field. Clearly the super-Alfv\'enic turbulence is full of a mixture of correlations as volume-filling tangled fields interact with networks of shocked gas, rarefactions and voids. Finally, we note that all of the exact values for the correlations discussed in this section are sensitive to the driving routine and numerical methods, as highlighted by \citet{Yoon2016_pressure_balance}, but regardless of the numerical treatment, and the precise correlations, the most important point is that not only can we can treat $s$ and $\lB$ independently in the the sub-Alfv\'enic regime, but $p(\lB)$ is fundamentally a delta distribution centered at zero. We will use both of these important results in the next section. 
    
    \begin{figure}
        \centering
        \includegraphics[width=\linewidth]{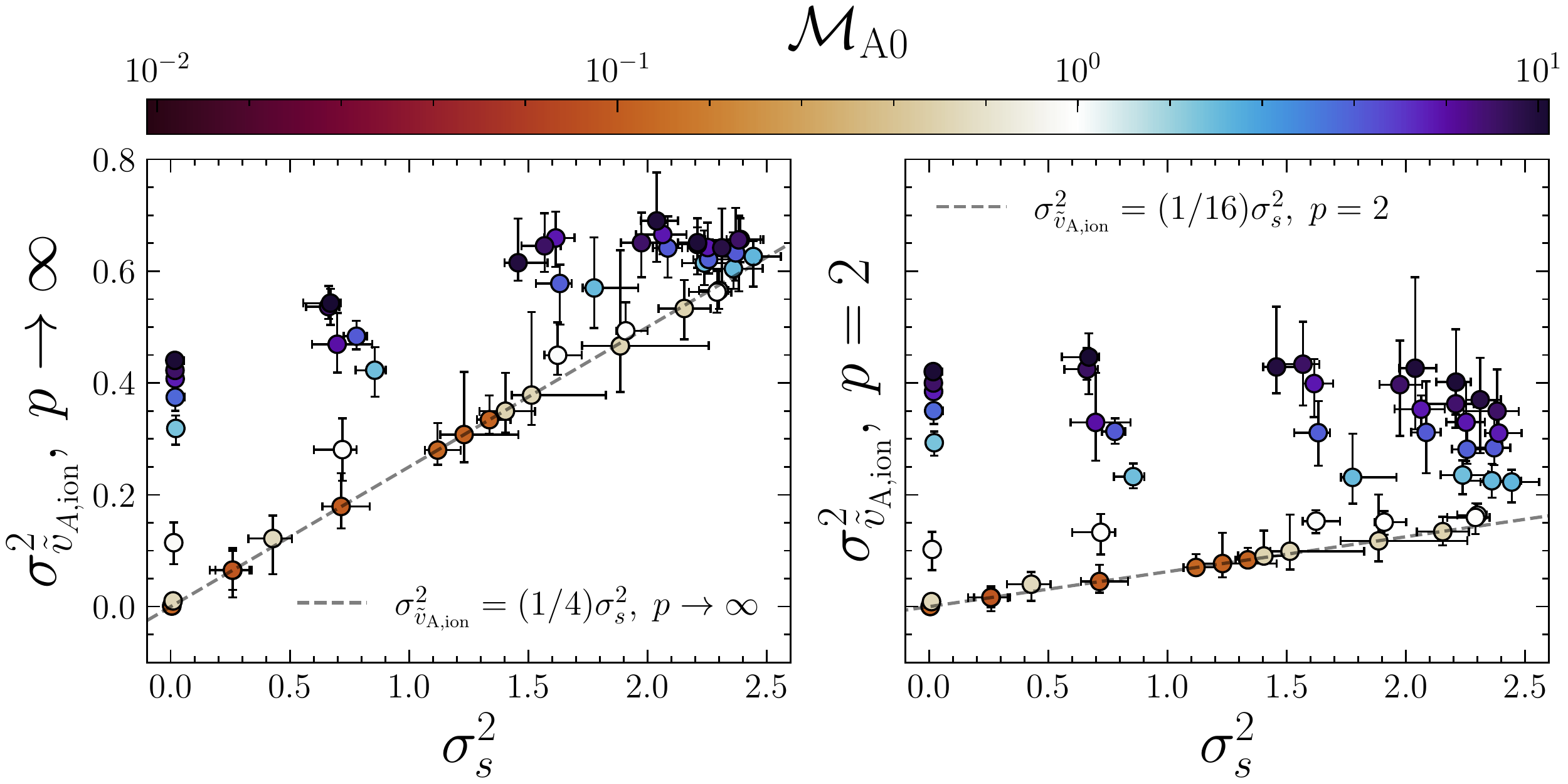}
        \caption{The logarithmic ion Alfv\'en velocity variance, $\var{\lvai}$, as a function of logarithmic density variance, $\var{s}$, colored by $\Mao$. The color bar diverges around the sub-to-super Alfv\'enic transition, with $\Mao < 1$ colored red and $\Mao > 1$ colored blue., \textbf{Left:} the variance for the ionization state when the gas density is evolving on short enough timescales to not have enough time to reach ionization equilibrium ($p\rightarrow\infty$). \textbf{Right:} the variance for when the gas is in ionization equilibrium ($p=2$). The $\var{\lvai}$ model (\autoref{eq:alfven_var_limit_p_inf} and \autoref{eq:alfven_var_limit_p_inf}) is shown with a gray-dashed line, and indicates the case where the ion Alfv\'en velocity fluctuations are controlled completely by density fluctuations, and the density and magnetic field fluctuations are independent from one another. In the sub-to-trans-Alfv\'enic regime, where these conditions are met, this model is a good description of the data, across all of $\var{s}$, and then for high-$\var{s}$ (high-$\M$) it gives a reasonable approximation, where the density fluctuations are strong. This highlights the importance of compressibility for understanding the ionic Alfv\'en velocity fluctuations.}
        \label{fig:va_rho_var}
    \end{figure}    
    
    \subsection{Constructing an ionic Alfv\'en velocity fluctuation model}\label{sec:model}
    Throughout the last two sections we have learned that the magnetic field fluctuations are extremely weak in sub-Alfv\'enic large-scale field turbulence and in fact negligible compared to the total power in the density fluctuations, as shown in the left panel of \autoref{fig:cov_Bs}. Furthermore, in this regime channel flows along $\Bo$ are the only way to significantly shock the gas, which makes $\var{\lB}$ and $\var{s}$ independent of one another. These two observations mean \autoref{eq:va_joint} becomes simply
    \begin{align} \label{eq:take_the_limit}
        \sigma_{\lvai}^2 = 
        \left(\frac{p-1}{2p}\right)^2\sigma_s^2, \text{ if } \Mao \leq 1,
    \end{align}  
    when it is completely determined by logarithmic density fluctuations.
    
    We plot $\var{\lvai}$ as a function of $\var{s}$ in \autoref{fig:va_rho_var} for both the $\chi = \chi_0 = 1$ case $(p\rightarrow\infty)$ and the $\chi \propto \rho^{-1/2}$, $(p=2)$ case, in the left and right panel, respectively. In the limit that all of the variation in $\lvai$ is explained by $s$-fluctuations, e.g., as appropriate for the sub-Alfv\'enic turbulent regime, \autoref{eq:take_the_limit} gives,
    \begin{align}\label{eq:alfven_var_limit_p_inf}
        \lim_{p \rightarrow \infty} \sigma_{\lvai}^2 = \frac{1}{4}\sigma_s^2, \text{ if } \Mao \leq 1,
    \end{align}
    and 
    \begin{align}\label{eq:alfven_var_limit_p_2}
        \sigma_{\lvai}^2(p=2) = \frac{1}{16}\sigma_s^2, \text{ if } \Mao \leq 1,
    \end{align}
    which we plot with the dashed-gray lines in each panel of \autoref{fig:va_rho_var}. For both ionization states, as expected from our analysis in \autoref{sec:magnetic} and above, the sub-to-trans-Alfv\'enic simulations closely match the relation, showing that indeed the variances of the ion Alfv\'en speeds are being controlled by the density fluctuations\footnote{Note that this means that the energy spectrum of Alfv\'en velocities must therefore also be controlled by the density. This is a very interesting repercussion of this result, and clearly demonstrates a stark difference between incompressible MHD turbulence, which always has Alfv\'en velocities being determined by magnetic field fluctuations, and compressible MHD turbulence. We leave the in-depth study of the two-point statistics, such as the structure functions and power spectra for future works.}. In the $p\rightarrow\infty$ state, the super-Alfv\'enic experiments at higher $\sigma_s^2$ (higher $\M$) start approaching the relation, even though the correlations between $s$ and $\lB$ are becoming significant (right panel of \autoref{fig:cov_Bs}). This is where the $s$-fluctuations are the strongest (the gas density is extremely inhomogenous) \citep[e.g., ][]{Molina2012_dens_var,Beattie2021_multishock} and hence still significantly contribute to the dispersion of the Alfv\'en velocities.
    
    \begin{figure*}
        \centering
        \includegraphics[width=\linewidth]{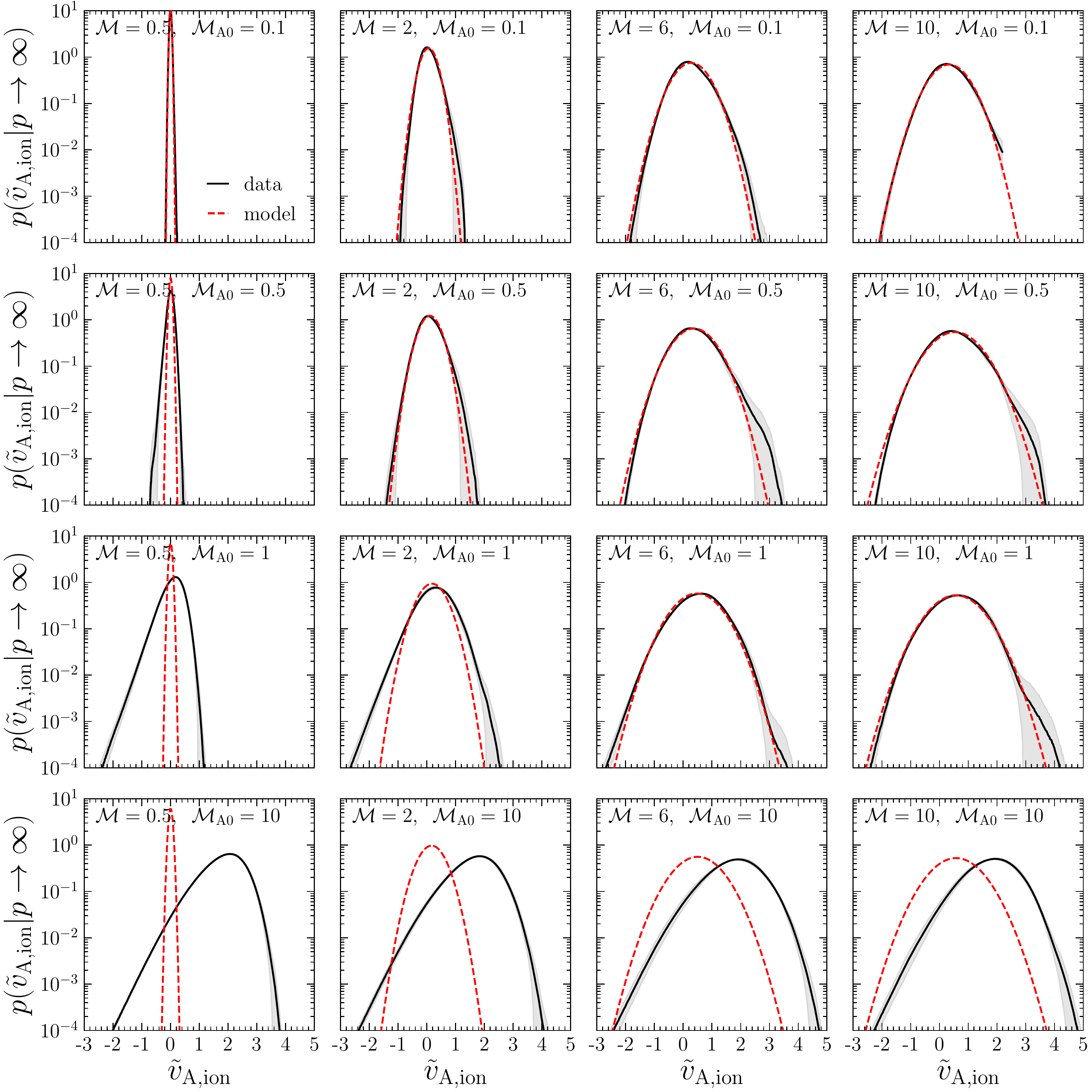}
        \caption{The volume-weighted logarithmic ion Alfv\'en velocity distribution for a selection of subsonic (left), supersonic (right) and sub-to-trans-Alfv\'enic (first three rows) simulations, shown in black, with $1\sigma$ fluctuations shown with the gray band. Our model, \autoref{eq:va_pdf_model}, is shown with the red dashed line, evaluated in the $\chi = \chi_0 = 1$ limit, corresponding to $p \rightarrow \infty$ in \autoref{eq:va_pdf_model_p}. $\Mao = 10.0$ simulations are shown in the last row, where our model is predicted to not be valid. The deviations between our model and the data in these simulations showcases the affect of magnetic field fluctuations and correlations on the $\lva$-PDF.}
        \label{fig:va_pdf}
    \end{figure*}
    
    Now that we have a variance model we move on to constructing the full $\lvai$-PDF, based on our previous results. Because $s$ and $\lB$ are independent in the $\Mao \lesssim 1$ regime, and $\lB$ is approximately a delta distribution centered at the zero, $\delta(\lB)$, it follows that,
    \begin{align}
        p(\lvai | \var{s}, p) = p\left(-s\frac{p-1}{2p}\right)\otimes p(\lB) = \int \d{\lB}\, p\left(-s\frac{p-1}{2p}\right) \delta(\lB) = p\left(-s\frac{p-1}{2p}\right),
    \end{align}
    which, for a lognormal $s$-PDF (a reasonable approximation, as discussed in \autoref{sec:dens}) $\lvai$ also becomes lognormal with $-(p-1)/(2p)$ propagated through the moments\footnote{Note $\Exp{aX}=a\Exp{X}$ and $\text{Var}\left\{aX\right\} = a^2\text{Var}\left\{X\right\}$, for random variable $X$ and constant $a$.}. It is
    \newpage
    \begin{align}
        p(\lvai | \var{s}, p ) &= \sqrt{\frac{2p}{\pi\sigma_s^2(p-1)}}\exp\left\{ - \frac{\left[\lvai - \var{s}\left(\frac{p-1}{2p}\right)^2\right]^2}{\sigma_s^2(p-1)/(2p)} \right\}, \label{eq:va_pdf_model_p}
    \end{align}        
    \begin{align}    
        \sigma_s^2 &= \left(\frac{2p}{p-1}\right)^2\sigma_{\lvai}^2, \label{eq:va_var_model_p}
    \end{align}
    and under the lognormal formalism, is solely dependent upon the $s$ variance, and ionization-density correlation $\chi \propto \rho^{-1/p}$. The strength of the $1/p$ correlation contracts (by a factor of $[2p/(p-1)]^{2}$) and shifts (by a factor of $-[2p/(p-1)]^{2}/2$, because $\Exp{\lva} = -[2p/(p-1)]^{2}/2\var{\lva}$) the $s$-PDF, mapping the logarithmic densities into the Alfv\'en velocities. As the correlation increases, $p \rightarrow 0$, the distribution approaches a delta distribution centered at zero, corresponding to no variation in the Alfv\'en velocities. 
    
    To compare with our data we take the limit in which $p \rightarrow \infty$. This reduces \autoref{eq:va_pdf_model_p} and \autoref{eq:va_var_model_p} to
    \begin{align}
        \lim_{p \rightarrow \infty} p(\lvai | \sigma_s^2, p ) &= \sqrt{\frac{2}{\pi\sigma_s^2}}\exp\left\{ - \frac{(\lvai - \sigma_s^2/4)^2}{\sigma_s^2/2} \right\}, \label{eq:va_pdf_model}
    \end{align}  
    \begin{align}    
        \lim_{p \rightarrow \infty} \sigma_s^2 &= 4\sigma_{\lvai}^2. \label{eq:va_var_model}
    \end{align}    
    In \autoref{fig:va_pdf} we show fits of \autoref{eq:va_pdf_model} for a representative sample of $\M$ and for the sub-to-trans-Alfv\'enic regime in the first three rows, and $\Mao = 10$ simulations in the final row, where we expect our model to be invalid. There is no fitting parameters in these PDF models, and they are solely determined by $\sigma_s^2$, which is measured independently in the data and fed into \autoref{eq:va_pdf_model}. 
    
    In the sub-to-trans-Alfv\'enic regime, the most significant deviation is at small $\M$ (low-$\var{s}$ in \autoref{fig:va_rho_var}). This is because, as shown in both panels in \autoref{fig:cov_Bs}, the subsonic, trans-Alfv\'enic regime has stronger magnetic field fluctuations than density fluctuations. These broaden the $\lva$ more than predicted in \autoref{eq:va_pdf_model} and imprint some of the negative skewness that we saw in \autoref{fig:b_pdf} on the $\lva$-PDF. Note also that for $\Mao = 1$, the $\M = 0.5$ experiment has the strongest covariance, as shown in \autoref{fig:cov_Bs}, so there are also density and magnetic field correlations developing in the plasma that we do not account for in this PDF model. For the rest of the sub-to-trans-Alfv\'enic simulations, all of the models describe the data well, with slight underestimates (not significant at 1$\sigma$, shown with the gray bands) of the high-$\lva$ tail. These might be due to $B/B_0 > 1$ regions in the plasma, where the Alfv\'en wave speeds can become very high. However, since the magnetic field fluctuations are very small, these are never very significant in this regime.
    
    We also include $\Mao = 10$ simulations in the bottom row. Clearly, based on the results discussed in \autoref{sec:dens} and \autoref{sec:magnetic}, \autoref{eq:alfven_var_limit_p_inf} is not valid in this regime. However, we still are able to learn about the plasma when it is in this state. Similar to the sub-to-trans-Alfv\'enic experiments, the low-$\M$ simulations are the dominated by magnetic field fluctuations, and are described poorly by only the density fluctuations. However, variance of the $\lvai$-PDFs for $\M > 2$ look reasonable (as also shown in the left panel of \autoref{fig:va_rho_var}) and it is just the mean that is not properly captured by the model. From \autoref{fig:b_pdf} and discussed in \autoref{sec:magnetic}, we know that the mean in the $\lB$-PDFs shifts with $\Mao$ and not $\M$. This means it is likely that the shift in the $\lvai$-PDFs is from the tangled magnetic field growing the Alfv\'en velocities everywhere in the plasma, even though the variance is still being controlled largely by the density inhomogeneities.
    
    We have now created a model for the PDF and variance of the logarithm of the Alfv\'en velocities that works over a broad range of $\M$ in the sub-to-trans-Alfv\'enic, relevant to interstellar medium turbulence. Next we discuss the repercussions of this model for using the $s$-PDF, which is an observational quantity, to measure $\lvai$-PDF, and cosmic ray propagation. 
    
\section{Discussion \& implications}\label{sec:discussion}

    \subsection{Measuring the $\lvai$-PDF}\label{sec:dis:pdf}
    
    Understanding the $\lvai$-PDF can directly help understand the streaming speeds of SCRs in local regions of the ISM. The $s$-PDF, which we utilized to construct the $\lvai$-PDF, is an accessible connection between theory and observations in the ISM, utilizing dust continuum emission \citep{Rathborne2015,Federrath2016_brick,Beattie2019b} or tracers such as $^{12}$CO, $^{13}$CO, or C$^{18}$O, J$=2-1$ emission \citep{Menon2020b,Sharda2021_driving_mode}. However, since the tracers of the gas density are projected onto the field of view (the column density), one has to connect the column density, $\Sigma = \int\d{\ell}\, \rho$, where $\ell$ is a line-of-sight length scale, to the three-dimensional volume gas density, $\rho$. One method of performing such a transformation is outlined in \citet{Brunt2010a}. \citet{Brunt2010a,Brunt2010b} used a projection-slice theorem in $k$-space, relating column to slices of the 3D gas density, and then applied a 2D-3D correction, $\mathcal{R}$\footnote{$\mathcal{R}$ is simply the ratio between the total integrated power in a power spectra in 2D (e.g., a column density map) versus 3D (e.g., the volumetric gas density), assuming perfect rotational symmetry (isotropy).}, assuming isotropy, such that $\var{\rho/\rho_0} = \mathcal{R} \sigma_{\Sigma / \Sigma_0}^2$. Furthermore, assuming a lognormal $\rho/\rho_0$-PDF, one can directly access the $s$ variance using $\sigma_s^2 = \ln(1 + \sigma_{\rho/\rho_0}^2)$, which is becoming a commonly used methodology for accessing $\var{s}$ \citep[see, for example, ][]{Federrath2016_brick,Menon2020b,Sharda2021_driving_mode}. This means, one can construct the $\lvai$-PDF using the following steps:
    \begin{enumerate}
        \item choose a trans-to-sub-Alfv\'enic region of the ISM \citep{HuaBai2013,Federrath2016_brick,Hu2019,Heyer2020,Hwang2021,Hoang2021,Skalidis2021_obs_sub_alf},
        \item obtain the column density and measure $\Sigma/\Sigma_0$ and compute $\var{\Sigma/\Sigma_0}$,
        \item apply the \citet{Brunt2010a} correction factor to derive $\var{\rho/\rho_0}$,
        \item use  $\sigma_s^2 = \ln(1 + \sigma_{\rho/\rho_0}^2)$ to compute the logarithmic density variance,
        \item and finally, use \autoref{eq:va_var_model}, $\var{\lva} = [(p-1)/(2p)]^2 \var{s}$ to fully describe a lognormal model for $v_{A,\rm{ion}}/v_{A0\,\rm{ion}}$, as in \autoref{eq:va_pdf_model}.
    \end{enumerate}
    As discussed in \autoref{sec:ionization _state}, the choice of $p$ depends on the phase being observed. In the cold ISM, $p=2$ because ionization equilibrium is reached quickly, while for diffuse atomic gas $p \rightarrow \infty$ because ionization equilibrium is not attained before density fluctuations dissipate. We now turn to the physical process that our analysis of the $\vai$ variance itself will help better understand -- the macroscopic diffusion of cosmic rays undergoing the streaming instability in the highly-magnetized regions of the ISM.
    
    \subsection{Modeling the parallel macroscopic SCR diffusion}\label{sec:dis:transport}
    
    For the sub-Alfv\'enic regime, where the magnetic field lines are dominated by the large-scale, non-turbulent component (see discussion in \autoref{sec:intro:mag}), channel flows form along the field lines and give rise to the density fluctuations (see \autoref{fig:joint_b_s}), which in turn control the $v_{A,\rm{ion}}$ fluctuations. In this regime we expect much faster transport (both streaming and macroscopic diffusion) along field lines than across them, and we expect the amount of parallel macroscopic diffusion to be much more sensitive to the $\vai$ fluctuations than the turbulent velocities because $\vai \gg v$. In \autoref{sec:va_fluctuations}, we found that $\vai$ follows a lognormal distribution. This means that the diffusive process that the SCRs take along magnetic field lines cannot possibly be regular Gaussian diffusion (where step-sizes are drawn from a Gaussian distribution), with $\big<x_{\parallel}^2\big> \propto t$, where $\big<x_{\parallel}^2\big>$ is the dispersion along a field line. A similar phenomena (but on micro-physical scale) has been shown for CR transport, with measurements coming from numerous numerical experiments indicating that CR diffusion is superdiffusive, $\big<x_{\parallel}^2\big> \propto t^\alpha$ with $\alpha>1$ \citep{Xu2013,lazarian2014superdiffusion,litvinenko2014analytical,Hu2022_superdiffusion_CRs}. 
    
    The regular explanation for superdiffusion in turbulent fluids is associated with \citet{Richardson1926_diffusion} diffusion: turbulent advection of field lines causes them to separate at a rate $\big<x_{\perp}^2\big> \propto t^{3/2}$ \citep[see Appendix C6 in][]{Schekochihin2020_bias_review}. But, of course, when the large-scale field is strong, as is the case in $\Mao < 1$ turbulence, this can only explain the diffusive process perpendicular to the large-scale field. Certainly, an attractive explanation is that the superdiffusive nature of parallel SCR diffusion may be held in the lognormal density structure of the turbulence, but this is speculation since it is not clear how a lognormal step-size distribution in velocities even asymptotically influences the diffusion, other than certainly not being Gaussian.
    
    \begin{figure}
        \centering
        \includegraphics{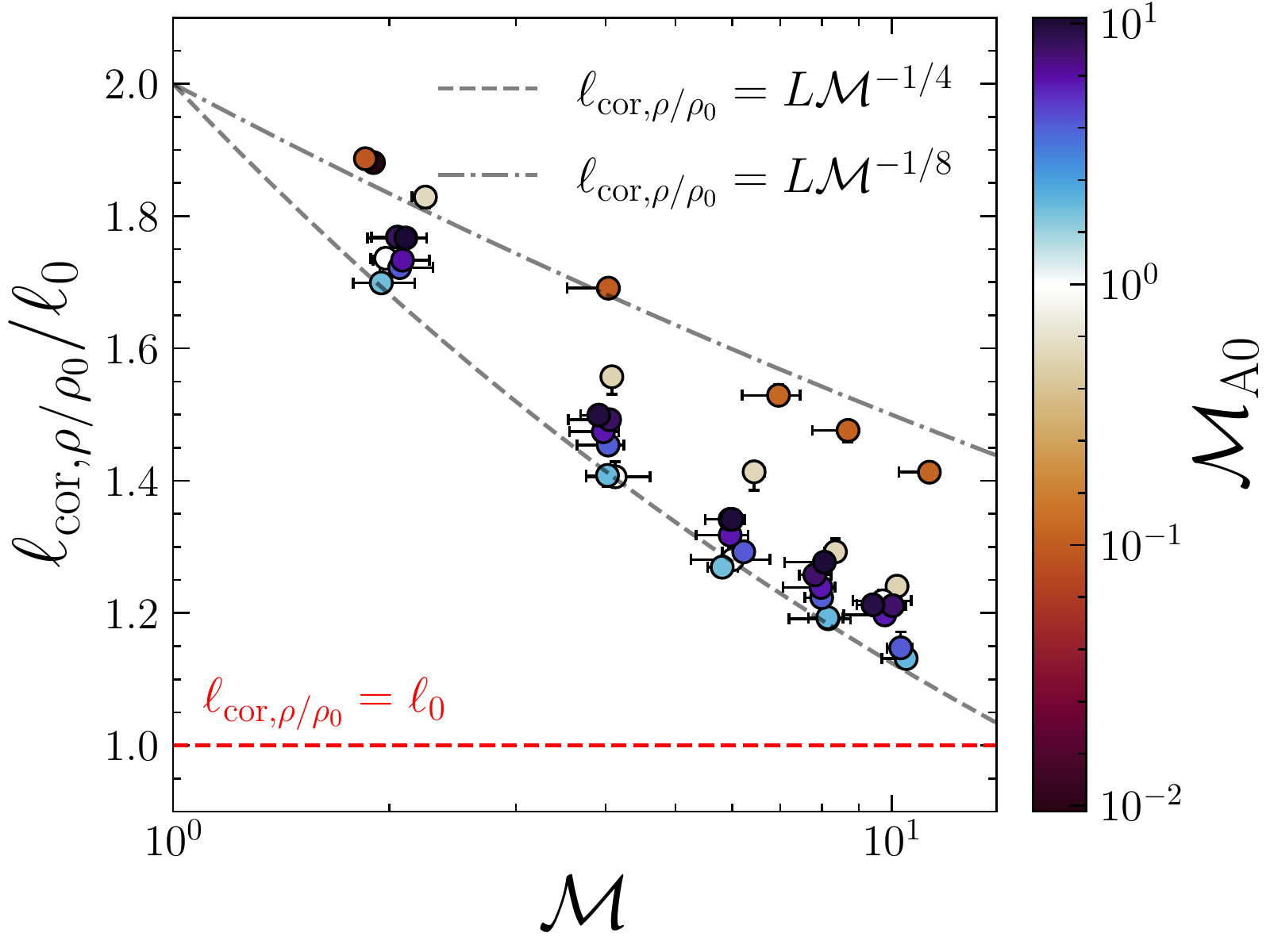}
        \caption{The correlation scale of $\rho/\rho_0$ for the supersonic experiments, relevant to our $\kapar$ model, \autoref{eq:diff_coef_model}, as computed with \autoref{eq:correlation_scale}, as a function of $\M$, colored in the same fashion as \autoref{fig:va_rho_var}. We show the driving scale, $\ell_0$, with the red horizontal line. We annotate an upper and lower power law envelope, $\cor{\rho/\rho_0} \propto \M^{-1/8} - \M^{-1/4}$, for the super-to-sub-Alfv\'enic experiments, respectively. In general, as $\M$ increases, the density fluctuations become correlated on smaller spatial scales, and the magnetic field suppresses the small-scale fluctuations.}
        \label{fig:dens_correlation_scale}
    \end{figure}    
    
    We leave a full exploration of the nature of parallel superdiffusion in these highly-magnetized plasmas for future work. However, to get at least an understanding of the magnitudes involved in how the density inhomogeneities enhance the along-field diffusion we can approximate the lognormal $\vai$ distribution as a Gaussian process. Assuming, $\sigma_{\lvai}<1$ (which is true for the sub-Alfv\'enic plasma, see \autoref{fig:va_rho_var}), the maximum error in the cumulative density functions for approximating a lognormal process as a Gaussian is the well known quantity, $\sigma_{\lvai}/(\sqrt{2\pi}e) + \var{\lvai}/(6\sqrt{\pi}e) \sim \mathcal{O}(10^{-3}-10^{-2})$ for $\Mao < 1$. This is small enough not to influence the total amplitude of the fluctuations, but may be important for the non-Gaussian effects related to the superdiffusion noted previously. Making this assumption, we can explicitly write a Markovian (Langevin-type) model \citep[e.g.,][]{Scannapieco2018,Mocz2019} for a SCR population moving along a field line in the $x_{\parallel}$ direction,
    \begin{align}\label{eq:markov_chain}
        x_{\parallel}(t + \d{t}) = x_{\parallel}(t) + \overbrace{\vaoi\d{t}}^{\text{advective term}} + \underbrace{\mathcal{N}(0,1)\sqrt{\frac{\var{\lvai}\ell_{\rm cor}^2}{t_{\rm cor}}\d{t}}}_{\text{diffusion term}},
    \end{align}
    which captures how the SCR populations are advected at the streaming speed, $\vaoi$ with some amount of macroscopic, Gaussian diffusion, or stochasticity through the Wiener process, $\mathcal{N}(0,1)\sqrt{\var{\lvai}\ell_{\rm cor}^2\d{t}/t_{\rm cor}}$, where $\mathcal{N}(0,1)$ is a standardized Gaussian distribution of mean zero. The amplitude of the fluctuations for the Wiener process is $\var{\lvai}\ell_{\rm cor}^2/t_{\rm cor}$, where $\ell_{\rm cor}^2/t_{\rm cor} = \vaoi \ell_{\rm cor}$, which is the expected size of a fluctuation for a SCR traveling at $\vaoi$, over the correlation scale $\ell_{\rm cor}$ during correlation time $t_{\rm cor}$. For the $\Mao < 1$ plasmas, the field lines are fundamentally completely straight, which in principle means that we simulate scales smaller than the correlation length of the magnetic field. Therefore, $\ell_{\rm cor}$ has to be the correlation scale of the density. This is approximately the correlation scale of the turbulence, but with some dependence upon $\M$ \citep[e.g., higher-$\M$ gives rise to more small-scale structure, which shifts the correlation scale to higher $k$ modes; ][]{Kim2005}. From \autoref{eq:markov_chain} one can immediately read-off the parallel macroscopic diffusion coefficient, 
    \begin{align} \label{eq:diff_coef_model}
        \kapar \approx \var{\lvai}\vaoi\ell_{\rm cor} = \left(\frac{p-1}{2p}\right)^2\var{s} \vaoi\ell_{\rm cor},
    \end{align}
    utilizing that $\var{\lvai} = ([p-1]/2p)^2\var{s}$ when $\Mao < 1$, as in \autoref{eq:take_the_limit}. Note that this is very similar to the parallel diffusion coefficient constructed by \citet{Krumholz2020} (see their Equation~21), but that the correlation length that appears here is the density correlation length, $\ell_{\rm cor} = \cor{\rho/\rho_0}$, rather than the magnetic field correlation length, since we have found that it is the density rather than magnetic field structure that facilitates the parallel diffusion in the sub-Alfv\'enic regime. Before using typical ISM parameters and computing $\kapar$, we first explore $\cor{\rho/\rho_0}$ in our numerical simulations.
    
    We directly compute the $\rho/\rho_0$ correlation scale using the textbook definition,
    \begin{align}\label{eq:correlation_scale}
        \frac{\cor{\rho/\rho_0}}{\ell_0} = \frac{L}{\ell_0}\frac{\displaystyle\int_0^{\infty} \d{k}\, k^{-1} \mathscr{P}_{\rho/\rho_0} (k)}{\displaystyle\int_0^{\infty} \d{k}\,\mathscr{P}_{\rho/\rho_0} (k)},
    \end{align}
    where $\mathscr{P}_{\rho/\rho_0}(k)$ is the 1D power spectrum of $\rho/\rho_0$, in \autoref{fig:dens_correlation_scale}. Qualitatively similar to what is found in hydrodynamical turbulence, as $\M$ increases, more small-scale fluctuations are introduced into the gas density, and the correlation scale moves towards smaller length scales, with some scatter set by $\Mao$. In general, $\cor{\rho/\rho_0} > \ell_0$. We interpret this as meaning that the density fluctuations, which may be perturbed from the driving scale modes, are able to grow and become correlated on larger scales than the velocity structure that created the fluctuation. We show some bounding power laws in gray, with the sub-Alfv\'enic experiments roughly following $\cor{\rho/\rho_0} = L \M^{-1/4}$, and the super-Alfv\'enic experiments following $\cor{\rho/\rho_0} = L \M^{-1/8}$. The sub-Alfv\'enic experiments have larger correlation scales than their super-Alfv\'enic counterparts, which, as discussed in \autoref{sec:dens}, is most likely due to the magnetic field suppressing small-scale fluctuations in the gas density. The key result is that for our model in \autoref{eq:diff_coef_model}, in the $\M \gtrsim 2$, $\Mao \leq 1$ regime, $\cor{\rho/\rho_0} \approx (1.8 - 1.2)\ell_0$.
    
    Now we have all of the ingredients for estimating $\kapar$ in a relevant astrophysical system. Because our model works in a $\Mao \lesssim 1$ and $\M > 1$ plasma state, we choose a typical, cold MC, where $T\sim 10\,\text{K}$ \citep[e.g., ][]{Wolfire1995_isothermal_ISM}, $\chi \sim 10^{-6}$ and $[p-1]/[2p] = 1/4$ (see \autoref{sec:ionization _state}), $\M \sim 10$ on the cloud scale $L \sim 10\,\text{pc}$, \citep{Schneider2013,Federrath2016_brick,Orkisz2017,Beattie2019b} and, for at least some MCs, $\Ma < 1 \implies \Mao < 1$ \citep{HuaBai2013,Federrath2016_brick,Hu2019,Heyer2020,Hwang2021,Hoang2021,Beattie2022_inprep_energybalance}. We take $\Mao = 1$, which for $\M \sim 10$, implies that $\vao = c_s \M \sim 2\,\text{kms}^{-1}$ for a $c_s = [ k_{\rm B} T / (\mu m_{\rm H})]^{1/2} \sim 0.2\,\text{km\,s}^{-1}$ for $\mu = 2.33$ (H$_2$ + He composition), and $k_{\rm B}$, the Boltzmann constant. This means $\vaoi = \vao / \sqrt{\chi} = 2 \times 10^{3}\,\text{km\,s}^{-1}$. Using \autoref{eq:2shockModel}, we can estimate $\sigma_s^2 = 2.25$. Because we are considering $\M$ on $L = 10\,\text{pc}$, we ought to think of the cloud as an isolated system, where the $k$ modes of the turbulence above the cloud-scale are ``removed". This means that $\ell_0 = L$, and therefore $\cor{\rho/\rho_0} = 1.2\ell_0 = 12\,\text{pc}$, as per \autoref{fig:dens_correlation_scale}. Populating \autoref{eq:diff_coef_model} with these typical parameter values for sub-Alfv\'enic, supersonic MCs, then we find the parallel macroscopic diffusion coefficient is
    \begin{align} \label{eq:k_par_model_eval}
        \kapar \approx 1\times10^{27} \left(\frac{[p-1]/[2p]}{0.25}\right)^2\bigg(\frac{\sigma_s}{1.5}\bigg)^2\left(\frac{\vao}{2\,\text{kms}^{-1}}\right)\bigg(\frac{\chi}{10^{-6}}\bigg)^{-\frac{1}{2}}\left(\frac{\cor{\rho/\rho_0}}{12\,\text{pc}}\right)\, \text{cm}^2\,\text{s}^{-1}.
    \end{align}
    Based on measurements of $\Ma$ $(= \Mao)$, utilizing the velocity gradient technique in \citet{Hu2019}, \autoref{eq:k_par_model_eval} should be a reasonable approximation for the parallel diffusion of SCRs in the five star-forming Gould Belt clouds: Taurus ($\Mao = 1.19 \pm 0.02$), Perseus A ($\Mao = 1.22 \pm 0.05$), L 1551 ($\Mao = 0.73 \pm 0.13$), Serpens ($\Mao = 0.98 \pm 0.08$) and NGC 1333 ($\Mao = 0.82 \pm 0.24$), which is a region in the Perseus MC. We also note that the value we derive from the density statistics in \autoref{eq:k_par_model_eval} gives a very reasonable value based on previous diffusion coefficient estimates for MCs in \citet{Owen2021_MC_diff_coefs} (see \S3.2.2.) and \citet{Xu_2022_damping_in_ISM} (see \S3.3), which fall between $10^{25} - 10^{30}\text{cm}^2\text{s}^{-1}$ in the $\text{GeV}$ energy range. However, we stress that our model is for \textit{macroscopic} diffusion coefficients that arise from inhomomgeneities in the MHD plasma. This means that the values need not be the same as the microscopic diffusion coefficients arising from CR scattering below the scale of isotropization in the pitch angle scattering distribution\footnote{Note that even though we are not reporting upon microscopic diffusion coefficients, there is no lack of measuring them in the literature \citep[e.g.,][for some recent measurements utilizing 1D, local MHD-PIC simulations see ]{Bai2022_PICsims_of_SCRs}.}. In any case, we look forward to genuine empirical determinations of the effective diffusion coefficient in molecular clouds. This may become possible with the Cherenkov Telescope Array \citep{Pedaletti2013}.
    
    Finally, we perform an estimate of what might be considered a galactic average for the parallel diffusion coefficient that comes from the density inhomomgeneities in a Milky Way-like galaxy. Like our toy molecular cloud model, we assume $\Mao = 1$, which translates to energy equipartition between a large-scale galactic field and the turbulent motions \citep[not an uncommon assumption, e.g., ][]{Beck_2013_Bfield_in_gal}. We use a velocity dispersion of $\sigma_V \sim 9.0\,\text{km\,s}^{-1}$, a typical value of the warm ionized medium (WIM) \citep[the volume-filling phase of the thick gaseous disc,][]{Boulares1990_CR_equipartition}, which means $\vao = 9.0\,\text{km\,s}^{-1}$ in energy equipartition. The ISM is probably trans-sonic on average \citep{Gaesnsler_2011_trans_ISM,Seta2021b}, so we take $\M = 2$ as our fiducial value (the lowest $\M$ we can use in our model), which gives $\sigma_s \approx 0.7$ based on \autoref{eq:2shockModel}. Furthermore, the WIM is completely ionized $\chi = 1$ and therefore the gas density and ionization fraction become independent ($p\rightarrow\infty$; see \autoref{sec:ionization _state}). We take the driving scale of the turbulence as $3\,\rm{kpc}$ \citep{Boulares1990_CR_equipartition} which, plausibly could come from coherent flows of gas in the galactic fountain region that extend out to $\rm{kpc}$ scales. Based on \autoref{fig:dens_correlation_scale} this implies that $\cor{\rho/\rho_0} = 5.1\,\rm{kpc}$. Propagating these values into \autoref{eq:k_par_model_eval} gives $\kapar \approx 3 \times 10^{27}\,\text{cm}^2\,\text{s}^{-1}$, in good agreement with the macroscopic diffusion coefficients modeled for the WIM in \citet{Xu_2022_damping_in_ISM} (based on magnetic field statistics), which are bounded between $10^{26}-10^{28}\,\text{cm}^2\,\text{s}^{-1}$ depending upon the energies. We note, however, that for these parameters our model starts to break down (and worse at smaller $\M$) due to the influence of magnetic field fluctuations, as shown in \autoref{fig:va_pdf} and we explicitly do not include velocity advection, which may also boost the diffusion coefficient. Hence $\kapar$ is a lower bound of the diffusion, which is both intrinsically Gaussian and based on just the density statistics. In a forthcoming paper \citep{Sampson2022_inprep_SCR_diffusion} we provide a better estimate calibrated by numerical simulations including these effects.

    \subsection{Caveats in our study}
    This work has been done in the context of an isothermal equation of state, and it is well known that the ISM is a multiphase plasma \citep{Ferriere2001,Hennebelle2012,Seta2022_multiphase_dynamo}. However, any one of the stable phases is approximately isothermal \citep{Wolfire1995_isothermal_ISM,Omukai2005_isothermal_ism}. This means the results in our study are only applicable to cosmic ray transport within any single phase of the ISM. In our study we use a mixture (50:50 in energy) of isotropic compressible and solenoidal modes to establish and maintain the turbulence. As highlighted in \citet{Yoon2016_pressure_balance}, the driving prescription changes the nature of density and magnetic field correlations in the turbulence and more compressive driving will give rise to stronger density fluctuations (changing the $b$ parameter in \autoref{eq:2shockModel}) and intermittent events, whereas solenoidal driving will have the opposite effect \citep{Federrath2008,Federrath2009,Federrath2010,Konstandin2012}. However, in the sub-Alfv\'enic regime, we do not believe that the correlations (or lack thereof) that we established in \autoref{sec:correlations} will change nor the overall conclusion we make in this study. This is because it is the strong $\Bo$ that restricts the magnetic field fluctuations from ever becoming dominant, and with isotropic driving shocked gas will form along $\Bo$, inevitably facilitating the same uncorrelated joint PDF that we find in the left panel of \autoref{fig:joint_b_s}. This means that the density fluctuations, even if they change in magnitude for different turbulent driving, will always control $\var{\va}$ in the $\Mao < 1$ regime. The effect of changing the degree of compressibility in the driving would therefore simply be to (slightly) change the relationship between $\var{\va}$ and $\mathcal{M}$.
    
    In this work we utilize ideal MHD models, free of an explicit form for the strain rate tensor in the momentum equation, or resistivity in the induction equation. Hence, the dissipation in our turbulence is purely numerical. Because the ion Alfv\'en velocity fluctuations are dominated by the low-$k$ modes (see \autoref{fig:convergence}, which shows that the rms statistics converge quickly as the number of grid elements in the simulations increase) the macroscopic diffusion of SCRs ought to be also controlled by the low-$k$ modes (low in the case of observations may either correspond to modes comparable to the scale of the driving source, or the largest modes in the observational region that is being examined, see e.g., \citealt{Federrath2016_brick,Stewart2022_velocity_dispersion_technique}). This means the exact prescription for dissipation ought not to matter for the ion Alfv\'en velocity statistics that we describe in this study. Relevant to observations, as long as we are able to analyze approximately isothermal regions of the ISM that are not dominated by dissipation (e.g. turbulent regions), our results ought to provide some insight into the rms statistics and 1-point statistics on those scales.
    
    The turbulence damping processes are important for the microphysics of the streaming instability. Throughout \autoref{sec:dis:transport} we have assumed that the growth of the resonant hydromagnetic modes are balanced by the ion neutral damping rate (\citealt{Kulsrud1969_streaming}; and shown recently in PIC simulations, \citealt{Bai2022_PICsims_of_SCRs}), giving rise to $v_{\rm stream} \sim \vai$. However, the balance is sensitive to the physics of the damping process. For example, \citet{Plotnikov2021_faster_streamingspeeds} showed that when the ion neutral damping rate is fast compared to the growth rate of the hydromagnetic modes, streaming velocities can reach up to $\sim 10\vai$ (see the lower panel in Figure~12). This is simple to propagate into our diffusion coefficient model, \autoref{eq:diff_coef_model}, which becomes $\kapar \approx \var{\lvai}\alpha\vaoi\ell_{\rm cor}$, where $\alpha = 10$ for the situation where the streaming speed is $10\vai$. The net result is obviously a linear response in diffusion by the factor that $v_{\rm stream}$ is above $\vai$. Of course, in general, the models in \autoref{sec:dis:transport} will only be valid when the streaming instability mechanism is the source of CR transport, but the models for the rms statistics and PDFs of the Alfv\'en ion velocities should be valid more generally in compressible turbulent plasmas.
    
    In our MHD models, we also omit self-gravity. Collapsing regions excite turbulent modes \citep{Federrath2011_jeans_criterion,Higashi2021_turbulent_modes_in_collapse}, create power law structure (one or two separate power laws) in the high-density tails of the $s$-PDF \citep[e.g.,][]{Federrath2012,Federrath2013b,Burkhart2018,Jaupart2020_powerlaw_s_PDF,Khuller2021}, make the correlation scale of the density move to much smaller scales \citep{Federrath2013b}, and correlate the magnetic field and density based on the geometry of the collapse \citep[e.g.,][]{Tritsis2015,Mocz2018}. The analysis in this study thus corresponds to ``subcritical" regions of the ISM, where the turbulent kinetic energy is greater than the gravitational potential energy, which may correspond to a large fraction of the MCs in the Milky Way \citep[and simulated analogues][]{Dobbs2011_MCs_not_bound,Tress2020_unbound_MCs}. 
    
    Our macroscopic diffusion coefficient modeling in \autoref{sec:dis:transport} paints a simple picture for the diffusion of SCRs in a sub-Alfv\'enic turbulent medium: cosmic rays stream through correlation lengths of the gas density at a streaming velocity set by the large-scale magnetic field strength, modulated by the variance of the density field, which is a function of the turbulence and the large-scale magnetic field. This process results in dispersion of the displacements for the SCRs increasing by a variance every correlation scale. Clearly, this simple picture does not take into account some important processes, such as the contribution from the microphysical diffusion, magnetic field fluctuations (which we show are sub-dominant), and the turbulent fluctuations in the velocity field. Regardless, our model provides a measure of the impact of the density fluctuations alone on the diffusion process, which we know from \autoref{sec:va_fluctuations} is the dominant mechanism for controlling the dispersion in Alfv\'en velocities in the low $\Mao$ limit. This effect is generally not captured in galaxy-scale simulations, which lack the resolution to capture the statistics of turbulent density fluctuations \citep[e.g., ][see their \S5.1.3, point viii]{Hopkins2021_cr_transport_models}. This means the effects that we are highlighting are unaccounted for, and cannot be captured by simply changing the streaming velocities. Rather, they need to be incorporated as an independent (macroscopic) term in the sub-grid CR diffusion recipes of more global simulations. These effects are potentially important, because as \citet{Hopkins2021_cr_transport_models} finds, microphysical self-confinement models need enhanced diffusion to match observed gamma-ray luminosities.
    %The turbulence that we model in this study, which is contributing to the macroscopic diffusion is unresolved turbulence in more global simulations like shown in \citet{Hopkins2021_cr_transport_models} (as discussed in point (viii) in \S5.1.3 of that study), so the effects that we are highlighting are unaccounted for. These effects are important because as \citet{Hopkins2021_cr_transport_models} points out, fluctuations in the ion Alfv\'en velocity statistics act to reduce the gamma ray luminosities, in turn, reducing the $\kappa$ that is required to be consistent with their data. These effects are not captured by simply changing the streaming velocities, but rather need to be incorporated as an independent (macroscopic) term in the sub-grid CR diffusion recipes of more global simulations. 
    We present such a sub-grid prescription in forthcoming work, \citet{Sampson2022_inprep_SCR_diffusion}.

\section{Conclusions and Summary}\label{sec:conclusion}
    Cosmic rays undergoing the streaming instability (streaming cosmic rays; SCRs) travel along magnetic field lines at the ionic Alfv\'en velocity, and hence the dispersion, or fluctuations in the Alfv\'en velocities act to effectively diffuse populations of SCRs. We explore the nature of these fluctuations using a large ensemble of three-dimensional isothermal magnetized, compressible (mostly supersonic) turbulence simulations, capturing a wide set of plasma parameters relevant to the interstellar medium of galaxies. The key result in this study is that when the large-scale field is sub-to-trans-Alfv\'enic, the magnetic field fluctuations are sub-dominant to the density fluctuations. This means the Alfv\'en velocity fluctuations, and likewise for the ionic Alfv\'en velocity fluctuations, are controlled by changes in the density, highlighting not only the role of compressibility in dispersing populations of SCRs, but also in determining the Alfv\'en velocity statistics in compressible MHD turbulence. We list further key results of the study below.
    \begin{itemize}
        \item In \autoref{sec:ionization _state} we estimate the ionization  equilibrium times and compare them to the typical timescales for density fluctuations in a turbulent medium. We show that the assumption of (instantaneous) ionization  equilibrium, which leads to $\chi \propto \rho^{-1/2}$, (\autoref{eq:ion_equilibrium}) is relevant for understanding ion Alfv\'en fluctuations in molecular gas $(\chi \sim 10^{-5})$, e.g., star-forming regions in the ISM. However, for diffuse atomic gas $(\chi \sim 10^{-3}-10^{-1})$, the equilibrium and gas density fluctuation timescales become comparable, and hence we can treat $\chi$ as approximately spatially constant. 
        \item In \autoref{sec:va_fluctuations} we show that the logarithmic Alfv\'en velocity magnitudes can be written as a sum of the logarithmic magnetic field and density magnitudes, \autoref{eq:va_joint}, and hence we study the variance and volume-weighted PDFs for the logarithmic gas density, $s$ (\autoref{sec:dens}) and the logarithmic magnetic field amplitude, $\lB$ (\autoref{sec:magnetic}). We show that for supersonic MHD turbulence the gas density is approximately lognormally distributed and the variance approximately follows the relations previously derived in the literature. We derive analytical models for the magnetic field variance, and the logarithmic magnetic field variance in the sub-to-trans-Alfv\'enic regime and show that the logarithmic magnetic field PDFs admit to significant non-Gaussian features that increase with $\Mao$. We attribute these to space-filling tangled fields that occupy a large part of the volume in the turbulence.
        \item We measure and discuss the covariance between $s$ and $\lB$ in \autoref{sec:correlations}, and show that trans-to-sub-Alfv\'enic turbulence exhibits only weak spatial correlations in \autoref{fig:cov_Bs} and \autoref{fig:joint_b_s}. We attribute this to channel flows that are orientated along magnetic field lines, compressing gas perpendicular to field lines without becoming correlated with the magnetic field. We also show that the density variance can be many orders of magnitude larger than the magnetic field variance in the sub-Alfv\'enic regime, and it is only in the subsonic, super-Alfv\'enic regime, where the magnetic field fluctuations have more total power than the density fluctuations. This implies that for $\Mao \lesssim 1$, and $\M > 1$, the dispersion in ion Alfv\'en velocities -- the speeds that Alfv\'en waves travel in compressible MHD -- are completely controlled by the density fluctuations. 
        \item Because the trans-to-sub-Alfv\'enic turbulence $\vai$ fluctuations are controlled by the density, which are approximately distributed lognormally, as discussed in \autoref{sec:intro:dens} and \autoref{sec:dens}, we are able to construct a lognormal $\vai$ theory, which we show in \autoref{sec:va_fluctuations}. In \autoref{fig:va_pdf} we show the PDF models, highlighting how they fit very well for the $\M>2$, low-$\Mao$ data, and break down at low-$\M$ and high-$\Mao$ because the magnetic field fluctuations become significant in either setting the dispersion, or shifting the $\lvai$ distribution to higher values than expected by purely density fluctuations. Based on this theory, we propose a method of determining the 3D-volume-weighted $\lvai$ PDF, and use it to determine an effective parallel diffusion parameter for populations of cosmic rays undergoing the streaming instability in a sub-Alfv\'enic plasma for a highly-magnetized molecular cloud environment and a lower bound for a Milky Way-like galactic average. 
    \end{itemize}
    
    \newpage
    \def\arraystretch{1.35}
\LTcapwidth=\textwidth
\begin{ThreePartTable}
\footnotesize\setlength{\tabcolsep}{5pt}
\begin{longtable}{l l}
\caption{Quantities and definitions used throughout this study}\\
\hline
\hline
Symbol \& Definition & Description  \\
\hline
$L$ & The length-scale of the system. \\[1em]
$\V = L^3$ & The volume of the system. \\[1em]
$\ell_0$ & The driving-scale of the turbulence. \\[1em]
$\displaystyle \frac{\cor{\rho/\rho_0}}{L} = \frac{\displaystyle\int_0^{\infty} \d{k}\, k^{-1} \mathscr{P}_{\rho/\rho_0} (k)}{\displaystyle\int_0^{\infty} \d{k}\,\mathscr{P}_{\rho/\rho_0} (k)}$ & The correlation, or integral scale of the gas density, where $\mathscr{P}_{\rho/\rho_0} (k)$ is the 1D power spectra. \\[1em]
$c_s$ & The sound speed. \\[1em]
$\displaystyle \M = \frac{\Exp{v^2}_{\V}^{1/2}}{c_s}$ & The sonic Mach number on $\ell_0$. \\[1em]
$\displaystyle \chi = \chi_0\left(\frac{\rho}{\rho_0}\right)^{1/p}$ & The ionization fraction by mass -  gas density correlation, where $\chi_0$ is the ionization fraction of the\\
& mean gas density, $\rho_0$, and $1/p$ is the correlation index. \\[1em]
$\displaystyle \va   = \frac{B}{\sqrt{4\pi\rho}}$ & The Alfv\'en velocity magnitude. \\[1em]
$\displaystyle \vao  = \frac{B_0}{\sqrt{4\pi\rho_0\chi}}$ & The Alfv\'en velocity magnitude of the mean-field quantities. \\[1em]
$\displaystyle \vai  = \frac{B}{\sqrt{4\pi\rho\chi}}$ & The ionic Alfv\'en velocity magnitude. \\[1em]
$\displaystyle \vaoi = \frac{B_0}{\sqrt{4\pi\rho_0\chi_0}}$ & The ionic Alfv\'en velocity magnitude of the mean-field quantities. \\[1em]
$\displaystyle \Ma = \frac{\Exp{v^2}_{\V}^{1/2}}{\va}$ & The Alfv\'en Mach number of the total field. \\[1em]
$\displaystyle \Mao = \frac{\Exp{v^2}_{\V}^{1/2}}{\vao}$ & The Alfv\'en Mach number of the mean-field quantities. \\[1em]
$\displaystyle \lB = \ln(B/B_0)$ & The logarithmic magnetic field normalized by the mean-field magnitude. \\[1em]
$\displaystyle \lva = \ln(\va/\vao)$ & The logarithmic Alfv\'en velocity normalized by the mean-field magnitude. \\[1em]
$\displaystyle \lvai = \ln(\vai/\vaoi)$ & The logarithmic ion Alfv\'en velocity normalized by the mean-field magnitude. \\[1em]
$\displaystyle s = \ln(\rho/\rho_0)$ & The logarithmic gas density normalized by the mean gas density. \\[1em]
$\kappa_{\parallel}$ & The macroscopic diffusion coefficient of streaming cosmic rays along the large-scale magnetic field \\
& due to inhomogeneities in $\vai$.  \\[1em]
\hline
\hline
\label{tb:symbol_glossary}
\end{longtable}
\end{ThreePartTable}
    \newpage
    \def\arraystretch{1.35}
\LTcapwidth=\textwidth
\begin{ThreePartTable}
\footnotesize\setlength{\tabcolsep}{5pt}
\begin{TableNotes}
\item \textbf{\textit{Notes:}} All simulations listed are run with grid resolutions of $16^3$, $36^3$, $72^3$, $144^3$ and $288^3$. All statistics are spatially averaged over the entire domain, $L^3 = \V$, and are computed for 51 time realizations, across 5 correlation times of the Ornstein-Uhlenbeck forcing function. From the distributions in time, we report the values for the $16^{\rm th}$, $50^{\rm th}$, and $84^{\rm th}$ percentiles. This process minimises the possibility of using statistics that are undergoing temporally intermittent turbulent events \citep{Beattie2021_spdf}. Column (1): the simulation ID, used throughout this study. Column (2): the turbulent Mach number, $\M \equiv \Exp{(\delta v / c_s)^2}^{1/2}_{\V}$. Column (3): the Alfv\'en Mach number of the mean magnetic field, $\Mao \equiv \Exp{ (\delta v \sqrt{4\pi\rho_0} ) / B_0}_{\V}$, with fluctuations coming from $\delta v$, since $\partial_{x_i}\mathbf{B}_0 = \partial_{t} \mathbf{B}_0 = 0$. Column (4): the mean magnetic field strength in units of $c_s \rho_0^{1/2}$. Column (5): the variance of the logarithm of $B/B_0$, $\lB \equiv \ln(B/B_0)$. Column (6): the same as column (5) but for the logarithm of densities, $s \equiv \ln(\rho/\rho_0)$. Column (7): the covariance between $\lB$ and $s$, $\sigma_{\lB, s} = \big<\big(\lB - \big<\lB\big>_{\V}\big)\big(s - \big<s\big>_{\V}\big)\big>_{\V}$. Column (8): the same as for column (6) but for the logarithmic Alfv\'en velocities, $\lva \equiv \ln(v_A/v_{A0})$. 
\end{TableNotes}
\begin{longtable}{l r@{}l r@{}l c r@{}l r@{}l r@{}l r@{}l}
\caption{Main simulation parameters and derived quantities used throughout this study.}\\
\hline
\hline
& \multicolumn{5}{c}{Simulation Parameters} & \multicolumn{8}{c}{Derived Quantities}\\
\hline
\multicolumn{1}{c}{Simulation ID} & \multicolumn{2}{c}{$\M$} & \multicolumn{2}{c}{$\Mao$} & \multicolumn{1}{c}{$\frac{B_0}{c_s \rho_0^{1/2}}$} & \multicolumn{2}{c}{$\var{\lB}$} & \multicolumn{2}{c}{$\var{s}$} & \multicolumn{2}{c}{$\sigma_{\lB,s}$} & \multicolumn{2}{c}{$\var{\lva}$}\\
\multicolumn{1}{c}{(1)} & \multicolumn{2}{c}{(2)} & \multicolumn{2}{c}{(3)} & \multicolumn{1}{c}{(4)} & \multicolumn{2}{c}{(5)} & \multicolumn{2}{c}{(6)} & \multicolumn{2}{c}{(7)} & \multicolumn{2}{c}{(8)} \\
\hline
\texttt{M05MA001}&$0.57$ &$_{-0.07}^{+0.02}$&$0.011$ &$_{-0.001}^{+0.0003}$&$180.0$&$5.5\ten{-9}$ &$_{-3\ten{-5}}^{+2\ten{-5}}$&$0.0045$ &$_{-0.009}^{+0.008}$&$-1.4\ten{-6}$ &$_{-2\ten{-7}}^{+2\ten{-7}}$&$0.0011$ &$_{-0.005}^{+0.004}$\\
\texttt{M05MA01}&$0.57$ &$_{-0.07}^{+0.02}$&$0.11$ &$_{-0.01}^{+0.004}$&$18.0$&$5.1\ten{-5}$ &$_{-0.003}^{+0.002}$&$0.0046$ &$_{-0.01}^{+0.01}$&$-9.9\ten{-5}$ &$_{-5\ten{-5}}^{+2\ten{-5}}$&$0.0013$ &$_{-0.005}^{+0.007}$\\
\texttt{M05MA05}&$0.53$ &$_{-0.04}^{+0.004}$&$0.53$ &$_{-0.04}^{+0.004}$&$3.5$&$0.009$ &$_{-0.02}^{+0.01}$&$0.01$ &$_{-0.02}^{+0.01}$&$0.0012$ &$_{-0.002}^{+0.002}$&$0.01$ &$_{-0.01}^{+0.009}$\\
\texttt{M05MA1}&$0.47$ &$_{-0.05}^{+0.01}$&$0.94$ &$_{-0.1}^{+0.03}$&$1.8$&$0.092$ &$_{-0.04}^{+0.03}$&$0.013$ &$_{-0.01}^{+0.01}$&$-0.019$ &$_{-0.003}^{+0.003}$&$0.11$ &$_{-0.04}^{+0.04}$\\
\texttt{M05MA2}&$0.46$ &$_{-0.03}^{+0.03}$&$1.8$ &$_{-0.1}^{+0.1}$&$0.89$&$0.27$ &$_{-0.02}^{+0.02}$&$0.021$ &$_{-0.03}^{+0.02}$&$-0.044$ &$_{-0.005}^{+0.007}$&$0.32$ &$_{-0.03}^{+0.02}$\\
\texttt{M05MA4}&$0.47$ &$_{-0.03}^{+0.06}$&$3.8$ &$_{-0.2}^{+0.5}$&$0.44$&$0.33$ &$_{-0.02}^{+0.03}$&$0.019$ &$_{-0.02}^{+0.04}$&$-0.044$ &$_{-0.006}^{+0.005}$&$0.37$ &$_{-0.02}^{+0.05}$\\
\texttt{M05MA6}&$0.48$ &$_{-0.03}^{+0.06}$&$5.8$ &$_{-0.4}^{+0.8}$&$0.3$&$0.36$ &$_{-0.01}^{+0.03}$&$0.019$ &$_{-0.02}^{+0.03}$&$-0.041$ &$_{-0.002}^{+0.003}$&$0.41$ &$_{-0.006}^{+0.03}$\\
\texttt{M05MA8}&$0.5$ &$_{-0.02}^{+0.06}$&$8.0$ &$_{-0.3}^{+1.0}$&$0.22$&$0.38$ &$_{-0.009}^{+0.01}$&$0.018$ &$_{-0.02}^{+0.02}$&$-0.036$ &$_{-0.003}^{+0.002}$&$0.42$ &$_{-0.007}^{+0.008}$\\
\texttt{M05MA10}&$0.51$ &$_{-0.03}^{+0.06}$&$10.0$ &$_{-0.5}^{+1.0}$&$0.18$&$0.4$ &$_{-0.02}^{+0.007}$&$0.017$ &$_{-0.01}^{+0.04}$&$-0.034$ &$_{-0.004}^{+0.003}$&$0.44$ &$_{-0.02}^{+0.01}$\\
\texttt{M05MA100}&$0.64$ &$_{-0.05}^{+0.09}$&$130.0$ &$_{-10.0}^{+20.0}$&$0.018$&$0.49$ &$_{-0.01}^{+0.02}$&$0.027$ &$_{-0.04}^{+0.05}$&$0.0016$ &$_{-0.01}^{+0.005}$&$0.5$ &$_{-0.01}^{+0.01}$\\
\texttt{M2MA001}&$1.9$ &$_{-0.06}^{+0.04}$&$0.0095$ &$_{-0.0003}^{+0.0002}$&$710.0$&$1.3\ten{-9}$ &$_{-6\ten{-6}}^{+4\ten{-5}}$&$0.26$ &$_{-0.1}^{+0.07}$&$-5.1\ten{-6}$ &$_{-1\ten{-6}}^{+6\ten{-7}}$&$0.065$ &$_{-0.05}^{+0.04}$\\
\texttt{M2MA01}&$1.9$ &$_{-0.02}^{+0.06}$&$0.093$ &$_{-0.001}^{+0.003}$&$71.0$&$1.3\ten{-5}$ &$_{-0.0005}^{+0.002}$&$0.26$ &$_{-0.07}^{+0.08}$&$-0.00046$ &$_{-0.0001}^{+8\ten{-5}}$&$0.065$ &$_{-0.04}^{+0.04}$\\
\texttt{M2MA05}&$2.2$ &$_{-0.09}^{+0.07}$&$0.56$ &$_{-0.02}^{+0.02}$&$14.0$&$0.01$ &$_{-0.02}^{+0.03}$&$0.43$ &$_{-0.1}^{+0.08}$&$-0.0044$ &$_{-0.006}^{+0.005}$&$0.12$ &$_{-0.06}^{+0.04}$\\
\texttt{M2MA1}&$2.0$ &$_{-0.09}^{+0.1}$&$0.99$ &$_{-0.05}^{+0.05}$&$7.1$&$0.068$ &$_{-0.02}^{+0.03}$&$0.72$ &$_{-0.1}^{+0.06}$&$-0.04$ &$_{-0.009}^{+0.01}$&$0.28$ &$_{-0.06}^{+0.06}$\\
\texttt{M2MA2}&$1.9$ &$_{-0.2}^{+0.2}$&$1.9$ &$_{-0.2}^{+0.2}$&$3.5$&$0.15$ &$_{-0.04}^{+0.02}$&$0.86$ &$_{-0.08}^{+0.05}$&$-0.064$ &$_{-0.02}^{+0.02}$&$0.42$ &$_{-0.05}^{+0.04}$\\
\texttt{M2MA4}&$2.1$ &$_{-0.1}^{+0.2}$&$4.1$ &$_{-0.3}^{+0.5}$&$1.8$&$0.23$ &$_{-0.05}^{+0.08}$&$0.78$ &$_{-0.05}^{+0.04}$&$-0.061$ &$_{-0.02}^{+0.07}$&$0.48$ &$_{-0.02}^{+0.03}$\\
\texttt{M2MA6}&$2.1$ &$_{-0.2}^{+0.2}$&$6.3$ &$_{-0.6}^{+0.6}$&$1.2$&$0.28$ &$_{-0.1}^{+0.1}$&$0.7$ &$_{-0.1}^{+0.1}$&$-0.011$ &$_{-0.04}^{+0.05}$&$0.47$ &$_{-0.05}^{+0.06}$\\
\texttt{M2MA8}&$2.1$ &$_{-0.2}^{+0.09}$&$8.2$ &$_{-0.7}^{+0.4}$&$0.89$&$0.39$ &$_{-0.03}^{+0.06}$&$0.66$ &$_{-0.09}^{+0.05}$&$0.019$ &$_{-0.03}^{+0.04}$&$0.54$ &$_{-0.02}^{+0.04}$\\
\texttt{M2MA10}&$2.1$ &$_{-0.2}^{+0.1}$&$11.0$ &$_{-1.0}^{+0.7}$&$0.71$&$0.43$ &$_{-0.1}^{+0.07}$&$0.67$ &$_{-0.1}^{+0.04}$&$0.057$ &$_{-0.06}^{+0.06}$&$0.54$ &$_{-0.04}^{+0.03}$\\
\texttt{M2MA100}&$2.4$ &$_{-0.2}^{+0.1}$&$120.0$ &$_{-8.0}^{+6.0}$&$0.071$&$0.79$ &$_{-0.08}^{+0.08}$&$0.72$ &$_{-0.1}^{+0.1}$&$0.41$ &$_{-0.08}^{+0.09}$&$0.57$ &$_{-0.02}^{+0.03}$\\
\texttt{M4MA01}&$4.0$ &$_{-0.5}^{+0.06}$&$0.1$ &$_{-0.01}^{+0.002}$&$140.0$&$2.8\ten{-5}$ &$_{-0.002}^{+0.0009}$&$0.71$ &$_{-0.08}^{+0.1}$&$-0.00036$ &$_{-0.0001}^{+0.0001}$&$0.18$ &$_{-0.04}^{+0.06}$\\
\texttt{M4MA05}&$4.1$ &$_{-0.1}^{+0.04}$&$0.51$ &$_{-0.01}^{+0.005}$&$28.0$&$0.0087$ &$_{-0.02}^{+0.02}$&$1.4$ &$_{-0.1}^{+0.1}$&$0.0086$ &$_{-0.006}^{+0.004}$&$0.35$ &$_{-0.04}^{+0.07}$\\
\texttt{M4MA1}&$4.1$ &$_{-0.4}^{+0.5}$&$1.0$ &$_{-0.09}^{+0.1}$&$14.0$&$0.058$ &$_{-0.04}^{+0.03}$&$1.6$ &$_{-0.06}^{+0.1}$&$0.0083$ &$_{-0.009}^{+0.02}$&$0.45$ &$_{-0.04}^{+0.06}$\\
\texttt{M4MA2}&$4.0$ &$_{-0.3}^{+0.2}$&$2.0$ &$_{-0.1}^{+0.08}$&$7.1$&$0.13$ &$_{-0.04}^{+0.06}$&$1.8$ &$_{-0.05}^{+0.2}$&$0.0064$ &$_{-0.02}^{+0.04}$&$0.57$ &$_{-0.07}^{+0.09}$\\
\texttt{M4MA4}&$4.0$ &$_{-0.4}^{+0.2}$&$4.0$ &$_{-0.4}^{+0.2}$&$3.5$&$0.25$ &$_{-0.1}^{+0.09}$&$1.6$ &$_{-0.1}^{+0.05}$&$0.08$ &$_{-0.04}^{+0.04}$&$0.58$ &$_{-0.07}^{+0.03}$\\
\texttt{M4MA6}&$4.0$ &$_{-0.4}^{+0.2}$&$6.0$ &$_{-0.6}^{+0.3}$&$2.4$&$0.34$ &$_{-0.06}^{+0.05}$&$1.6$ &$_{-0.05}^{+0.08}$&$0.078$ &$_{-0.04}^{+0.07}$&$0.66$ &$_{-0.05}^{+0.05}$\\
\texttt{M4MA8}&$4.1$ &$_{-0.5}^{+0.08}$&$8.1$ &$_{-1.0}^{+0.2}$&$1.8$&$0.41$ &$_{-0.1}^{+0.1}$&$1.6$ &$_{-0.1}^{+0.07}$&$0.13$ &$_{-0.05}^{+0.09}$&$0.65$ &$_{-0.05}^{+0.06}$\\
\texttt{M4MA10}&$3.9$ &$_{-0.2}^{+0.2}$&$9.8$ &$_{-0.6}^{+0.5}$&$1.4$&$0.43$ &$_{-0.06}^{+0.1}$&$1.5$ &$_{-0.06}^{+0.1}$&$0.17$ &$_{-0.03}^{+0.09}$&$0.62$ &$_{-0.03}^{+0.08}$\\
\texttt{M4MA100}&$4.2$ &$_{-0.08}^{+0.2}$&$110.0$ &$_{-2.0}^{+5.0}$&$0.14$&$1.1$ &$_{-0.04}^{+0.09}$&$1.6$ &$_{-0.2}^{+0.1}$&$0.88$ &$_{-0.1}^{+0.09}$&$0.65$ &$_{-0.03}^{+0.05}$\\
\texttt{M6MA01}&$7.0$ &$_{-0.8}^{+0.5}$&$0.12$ &$_{-0.01}^{+0.008}$&$210.0$&$5.2\ten{-5}$ &$_{-0.003}^{+0.0008}$&$1.1$ &$_{-0.05}^{+0.1}$&$3.1\ten{-5}$ &$_{-0.0002}^{+0.0002}$&$0.28$ &$_{-0.03}^{+0.05}$\\
\texttt{M6MA05}&$6.4$ &$_{-0.1}^{+0.2}$&$0.54$ &$_{-0.01}^{+0.02}$&$43.0$&$0.0086$ &$_{-0.01}^{+0.01}$&$1.5$ &$_{-0.08}^{+0.3}$&$0.015$ &$_{-0.005}^{+0.004}$&$0.38$ &$_{-0.05}^{+0.1}$\\
\texttt{M6MA1}&$6.0$ &$_{-0.8}^{+0.8}$&$1.0$ &$_{-0.1}^{+0.1}$&$21.0$&$0.044$ &$_{-0.03}^{+0.06}$&$1.9$ &$_{-0.04}^{+0.09}$&$0.027$ &$_{-0.02}^{+0.03}$&$0.49$ &$_{-0.03}^{+0.05}$\\
\texttt{M6MA2}&$5.8$ &$_{-0.3}^{+0.3}$&$1.9$ &$_{-0.09}^{+0.1}$&$11.0$&$0.13$ &$_{-0.03}^{+0.05}$&$2.2$ &$_{-0.09}^{+0.1}$&$0.066$ &$_{-0.03}^{+0.04}$&$0.61$ &$_{-0.04}^{+0.06}$\\
\texttt{M6MA4}&$6.2$ &$_{-0.4}^{+0.1}$&$4.2$ &$_{-0.3}^{+0.08}$&$5.3$&$0.24$ &$_{-0.08}^{+0.1}$&$2.1$ &$_{-0.06}^{+0.06}$&$0.13$ &$_{-0.04}^{+0.03}$&$0.64$ &$_{-0.05}^{+0.06}$\\
\texttt{M6MA6}&$6.0$ &$_{-0.6}^{+0.4}$&$6.0$ &$_{-0.6}^{+0.4}$&$3.5$&$0.29$ &$_{-0.05}^{+0.05}$&$2.1$ &$_{-0.1}^{+0.1}$&$0.15$ &$_{-0.04}^{+0.03}$&$0.67$ &$_{-0.04}^{+0.02}$\\
\texttt{M6MA8}&$6.0$ &$_{-0.5}^{+0.2}$&$7.9$ &$_{-0.6}^{+0.2}$&$2.7$&$0.39$ &$_{-0.1}^{+0.1}$&$2.0$ &$_{-0.09}^{+0.09}$&$0.24$ &$_{-0.06}^{+0.05}$&$0.65$ &$_{-0.06}^{+0.05}$\\
\texttt{M6MA10}&$6.0$ &$_{-0.5}^{+0.2}$&$10.0$ &$_{-0.8}^{+0.4}$&$2.1$&$0.41$ &$_{-0.1}^{+0.2}$&$2.0$ &$_{-0.07}^{+0.09}$&$0.22$ &$_{-0.04}^{+0.08}$&$0.69$ &$_{-0.07}^{+0.09}$\\
\texttt{M8MA01}&$8.7$ &$_{-0.9}^{+0.09}$&$0.11$ &$_{-0.01}^{+0.001}$&$280.0$&$3.7\ten{-5}$ &$_{-0.002}^{+0.0009}$&$1.2$ &$_{-0.1}^{+0.2}$&$-8\ten{-5}$ &$_{-0.0002}^{+0.0006}$&$0.31$ &$_{-0.05}^{+0.1}$\\
\texttt{M8MA05}&$8.4$ &$_{-0.3}^{+0.2}$&$0.52$ &$_{-0.02}^{+0.01}$&$57.0$&$0.0074$ &$_{-0.02}^{+0.01}$&$1.9$ &$_{-0.2}^{+0.4}$&$0.014$ &$_{-0.004}^{+0.009}$&$0.47$ &$_{-0.08}^{+0.2}$\\
\texttt{M8MA1}&$8.2$ &$_{-1.0}^{+0.6}$&$1.0$ &$_{-0.1}^{+0.08}$&$28.0$&$0.047$ &$_{-0.04}^{+0.02}$&$2.3$ &$_{-0.08}^{+0.04}$&$0.045$ &$_{-0.01}^{+0.02}$&$0.57$ &$_{-0.03}^{+0.04}$\\
\texttt{M8MA2}&$8.1$ &$_{-0.5}^{+0.4}$&$2.0$ &$_{-0.1}^{+0.1}$&$14.0$&$0.13$ &$_{-0.04}^{+0.05}$&$2.4$ &$_{-0.1}^{+0.1}$&$0.11$ &$_{-0.02}^{+0.03}$&$0.6$ &$_{-0.04}^{+0.06}$\\
\texttt{M8MA4}&$8.0$ &$_{-0.4}^{+0.3}$&$4.0$ &$_{-0.2}^{+0.1}$&$7.1$&$0.22$ &$_{-0.05}^{+0.07}$&$2.3$ &$_{-0.06}^{+0.09}$&$0.17$ &$_{-0.03}^{+0.03}$&$0.62$ &$_{-0.03}^{+0.04}$\\
\texttt{M8MA6}&$8.0$ &$_{-0.9}^{+0.4}$&$6.0$ &$_{-0.7}^{+0.3}$&$4.7$&$0.3$ &$_{-0.1}^{+0.09}$&$2.3$ &$_{-0.08}^{+0.08}$&$0.22$ &$_{-0.04}^{+0.03}$&$0.64$ &$_{-0.03}^{+0.04}$\\
\texttt{M8MA8}&$7.8$ &$_{-0.4}^{+0.4}$&$7.8$ &$_{-0.4}^{+0.4}$&$3.5$&$0.36$ &$_{-0.09}^{+0.07}$&$2.2$ &$_{-0.1}^{+0.07}$&$0.25$ &$_{-0.05}^{+0.08}$&$0.65$ &$_{-0.04}^{+0.03}$\\
\texttt{M8MA10}&$8.1$ &$_{-1.0}^{+0.2}$&$10.0$ &$_{-1.0}^{+0.3}$&$2.8$&$0.44$ &$_{-0.08}^{+0.08}$&$2.2$ &$_{-0.08}^{+0.06}$&$0.34$ &$_{-0.05}^{+0.06}$&$0.65$ &$_{-0.06}^{+0.04}$\\
\texttt{M10MA01}&$11.0$ &$_{-1.0}^{+0.3}$&$0.11$ &$_{-0.01}^{+0.003}$&$350.0$&$3.3\ten{-5}$ &$_{-0.003}^{+0.002}$&$1.3$ &$_{-0.05}^{+0.09}$&$6.3\ten{-5}$ &$_{-0.0004}^{+0.0001}$&$0.33$ &$_{-0.03}^{+0.04}$\\
\texttt{M10MA05}&$10.0$ &$_{-0.3}^{+0.3}$&$0.51$ &$_{-0.02}^{+0.01}$&$71.0$&$0.007$ &$_{-0.02}^{+0.007}$&$2.2$ &$_{-0.1}^{+0.1}$&$0.014$ &$_{-0.002}^{+0.005}$&$0.53$ &$_{-0.06}^{+0.05}$\\
\texttt{M10MA1}&$9.7$ &$_{-0.9}^{+0.9}$&$0.97$ &$_{-0.09}^{+0.09}$&$35.0$&$0.041$ &$_{-0.05}^{+0.04}$&$2.3$ &$_{-0.1}^{+0.06}$&$0.045$ &$_{-0.01}^{+0.02}$&$0.56$ &$_{-0.04}^{+0.04}$\\
\texttt{M10MA2}&$10.0$ &$_{-0.8}^{+0.3}$&$2.1$ &$_{-0.2}^{+0.07}$&$18.0$&$0.13$ &$_{-0.04}^{+0.05}$&$2.4$ &$_{-0.1}^{+0.1}$&$0.12$ &$_{-0.03}^{+0.02}$&$0.63$ &$_{-0.05}^{+0.03}$\\
\texttt{M10MA4}&$10.0$ &$_{-0.4}^{+0.4}$&$4.1$ &$_{-0.2}^{+0.2}$&$8.9$&$0.24$ &$_{-0.09}^{+0.08}$&$2.4$ &$_{-0.1}^{+0.07}$&$0.18$ &$_{-0.05}^{+0.05}$&$0.63$ &$_{-0.05}^{+0.08}$\\
\texttt{M10MA6}&$9.8$ &$_{-1.0}^{+0.3}$&$5.9$ &$_{-0.7}^{+0.2}$&$5.9$&$0.28$ &$_{-0.08}^{+0.05}$&$2.4$ &$_{-0.06}^{+0.1}$&$0.22$ &$_{-0.02}^{+0.04}$&$0.66$ &$_{-0.04}^{+0.04}$\\
\texttt{M10MA8}&$10.0$ &$_{-0.9}^{+0.4}$&$8.0$ &$_{-0.7}^{+0.3}$&$4.4$&$0.36$ &$_{-0.1}^{+0.07}$&$2.4$ &$_{-0.1}^{+0.09}$&$0.3$ &$_{-0.06}^{+0.08}$&$0.66$ &$_{-0.09}^{+0.04}$\\
\texttt{M10MA10}&$9.4$ &$_{-0.5}^{+0.2}$&$9.4$ &$_{-0.5}^{+0.2}$&$3.5$&$0.38$ &$_{-0.09}^{+0.07}$&$2.3$ &$_{-0.06}^{+0.09}$&$0.31$ &$_{-0.03}^{+0.02}$&$0.64$ &$_{-0.06}^{+0.07}$\\
\hline
\hline
\insertTableNotes
\label{tb:simtab}
\end{longtable}
\end{ThreePartTable}

\section*{Conflict of Interest Statement}
    The authors declare that the research was conducted in the absence of any commercial or financial relationships that could be construed as a potential conflict of interest.

\section*{Author Contributions}
    J.~R.~B. lead this study, performed the numerical simulations, produced the figures and wrote a majority of this manuscript. 
    M.~R.~K. wrote \autoref{sec:ionization _state}. 
    J.~R.~B., M.~R.~K., C.~F., M.~S. and R.~M.~C. contributed significantly to the development of the ideas presented in this study, and the detailed editing of the manuscript. 

\section*{Funding}
    J.~R.~B. acknowledges financial support from the Australian National University, via the Deakin PhD and Dean's Higher Degree Research (theoretical physics) Scholarships and the Australian Government via the Australian Government Research Training Program Fee-Offset Scholarship.
    C.~F. and J.~R.~B~acknowledge high-performance computing resources provided by the Leibniz Rechenzentrum and the Gauss Centre for Supercomputing (grants~pr32lo, pn73fi, and GCS Large-scale project~22542), and M.~R.~K., C.~F., and J.~R.~B. acknowledge high-performance computing resources provided by the Australian National Computational Infrastructure (grants~jh2 and~ek9) in the framework of the National Computational Merit Allocation Scheme and the ANU Merit Allocation Scheme.
    M.~R.~K. acknowledges support from the Australian Research Council's \textit{Discovery Projects} and \textit{Future Fellowship} schemes, awards DP190101258 and FT180100375.
    C.~F.~acknowledges funding provided by the Australian Research Council (Future Fellowship FT180100495), and the Australia-Germany Joint Research Cooperation Scheme (UA-DAAD).
    M.~L.~S. acknowledges financial support from the Australian Government via the Australian Government Research Training Program Stipend and Fee-Offset Scholarship.
    R.~M.~C. acknowledges support from the Australian Research Council's \textit{Discovery Project}, award DP190101258.

\section*{Acknowledgments}
    J.~R.~B.~thanks Christoph Federrath's and Mark Krumholz's research groups and Philip Mocz for many productive discussions. We thank the anonymous reviewers for their suggestions that helped enhance the clarity of this study.\\
    
    The fluid simulation software, \textsc{flash}, was in part developed by the Flash Centre for Computational Science at the Department of Physics and Astronomy of the University of Rochester. The turbulence driving module can be accessed from \citet{Federrath2022_turbulence_driving_module}. Data analysis and visualisation software used in this study: \textsc{C++} \citep{Stroustrup2013}, \textsc{numpy} \citep{Oliphant2006,numpy2020}, \textsc{matplotlib} \citep{Hunter2007}, \textsc{scipy} \citep{Virtanen2020} and \textsc{emcee} \citep{foreman2013emcee}.

\section*{Data Availability Statement}
    The data underlying this article will be shared on reasonable request to the corresponding author.

\section{Supplementary Tables and Figures}

\subsection{Parallel to large-scale field slice visualisation}
\begin{figure*}
        \centering
        \includegraphics[width=0.95\linewidth]{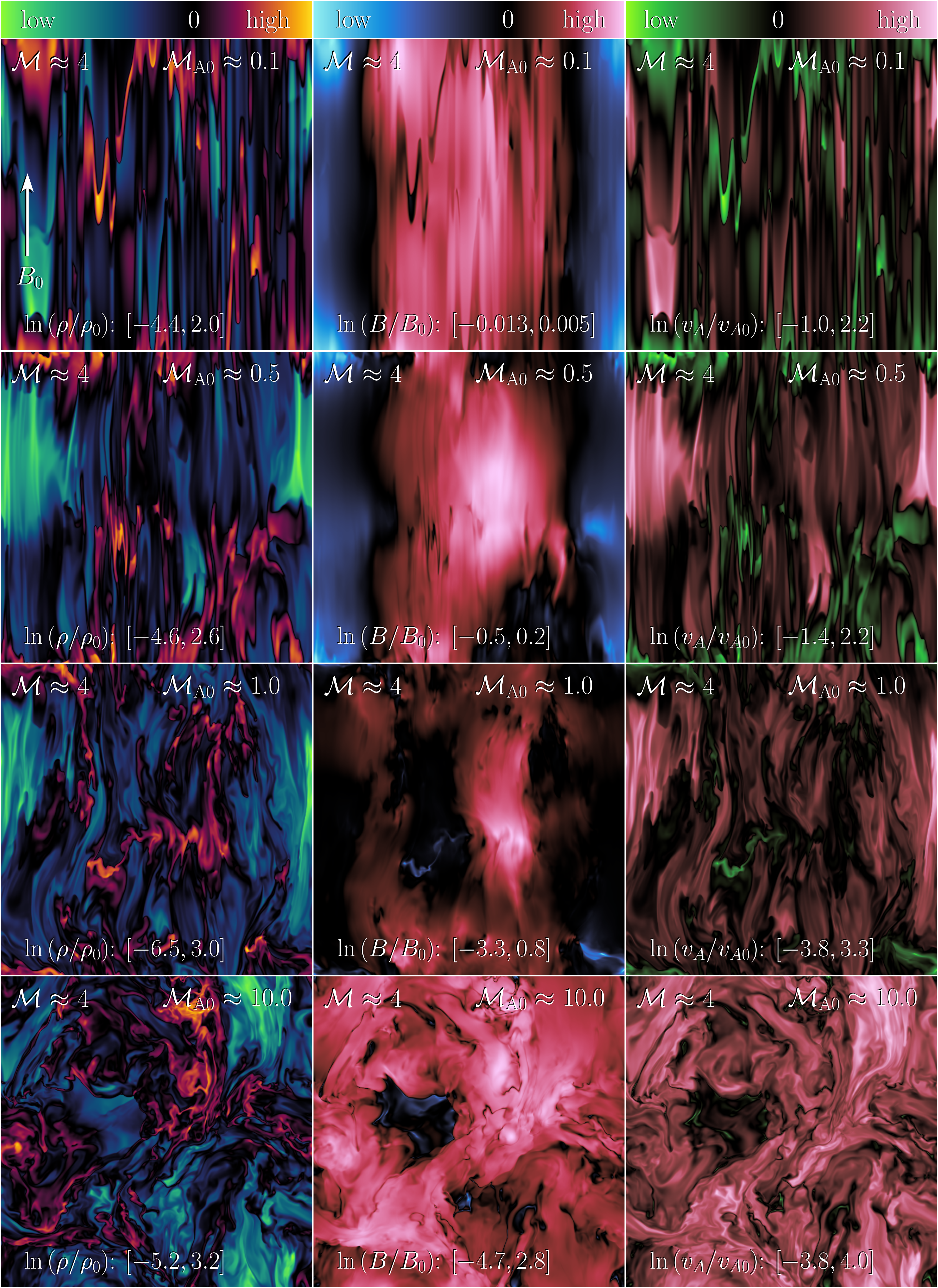}
        \caption{The same as \autoref{fig:12_panel_gas_vars} but for a slice parallel to the large-scale field revealing the fluctuations propagating up and down the field. The same correlations between field variables are present as in \autoref{fig:12_panel_gas_vars}.}
        \label{fig:12_panel_gas_vars_2}
    \end{figure*}

    In \autoref{fig:12_panel_gas_vars_2} we show a slice through the same dataset as in \autoref{fig:12_panel_gas_vars}, but for the slice plane parallel to the large-scale field. The direction of the field is shown in the upper-left panel of the plot. This slice orientation reveals the same correlation as in \autoref{fig:12_panel_gas_vars}, i.e., the gas density and Alfv\'en velocity in low-$\Mao$ simulations are strongly correlated with each other, and the magnetic field and Alfv\'en velocity are in the high-$\Mao$ simulations. More evident in this slice orientation is the formation of strong shocked gas traveling along the large-scale field, clearly responsible for creating strong fluctuations in both the density and Alfv\'en velocity at low-$\Mao$.

\subsection{Numerical Convergence}\label{sec:num_converge}

    \begin{figure}
        \centering
        \includegraphics[width=\linewidth]{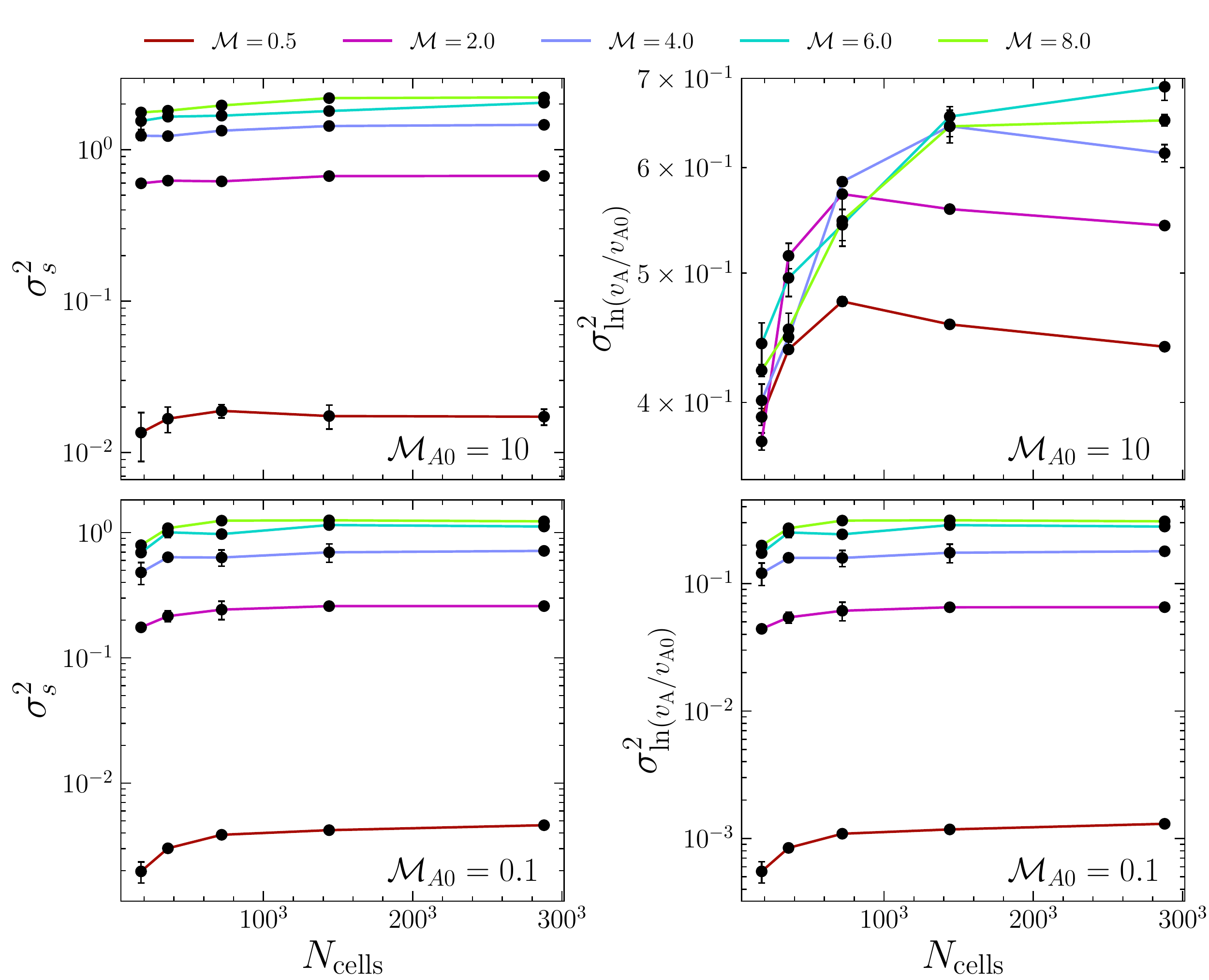}
        \caption{Numerical convergence tests for the two main statistics from our study. We show $\var{s}$ (left) and $\var{\ln(\va/\vao)}$ (right) as a function of the number of grid elements $N_{\rm res}$ that discretize the $L^3$ simulation domain. $N_{\rm res}$ varies from $18^3$ to $288^3$, for the ensemble of $\Mao = 10$ (top row) and $\Mao = 0.1$ (bottom row) simulations, colored by the different $\M$ so that one can distinguish between simulations, shown in the legend above the panels.}
        \label{fig:convergence}
    \end{figure}

    In \autoref{fig:convergence} we show numerical convergence tests for the two main statistics presented in this study: the variance of the logarithmic gas density (left), and the variance of the logarithmic Alfv\'en velocity magnitudes (right). For these tests, we compute the variances, averaged from $5\tau - 10\tau$, for $N_{\rm cells} = 18^3$, $36^3$, $72^3$, $144^3$ and $288^3$. In the top column we show the $\Mao=10$ simulation ensemble, and in the bottom panel the $\Mao=0.1$ panel, showcasing both super-and-sub-Alfv\'enic cases. At $288^3$, which is the grid resolution used in the main text of this study, all of the statistics are well-converged in grid resolution.

\newpage
\bibliographystyle{frontiersinSCNS_ENG_HUMS}
\bibliography{Jan2022}

\end{document}